\documentclass[aps,prd,11pt,superscriptaddress,notitlepage,longbibliography,nofootinbib,tightenlines,groupedaddress]{revtex4-1}
\usepackage{amsmath,amssymb,amsfonts}
\usepackage{graphicx}
\usepackage{color}
\usepackage{tikz}
\usepackage{dsfont}
\usepackage[pdftex]{hyperref}
\hypersetup{colorlinks=true, linkcolor=darkred, citecolor=blue, linktoc=page}
\definecolor{darkred}{rgb}{0.8,0.1,0.1}

\numberwithin{equation}{section}
\renewcommand\theequation{\arabic{section}.\arabic{equation}} %switch back for appendix

\usepackage{subfigure}

\def\cA{{\cal A}}

\def\cF{{\cal F}}

\def\cI{{\cal I}}
\def\RR{\ensuremath{\mathbb R}}

\def\ZZ{\ensuremath{\mathbb Z}}

\def\cL{{\cal L}}
\def\cW{{\cal W}}
\def\cZ{{\cal Z}}

\DeclareMathOperator{\sech}{sech}
\DeclareMathOperator{\csch}{csch}
\DeclareMathOperator{\tr}{tr}

\makeatletter\def\l@subsubsection#1#2{}%
\makeatother

\def\Im{\mathop{\rm Im}}
\def\Re{\mathop{\rm Re}}

\newcommand{\pslash}{\ensuremath\diagup\!\!\!\!\!{+}}

\begin{abstract}
Exact results are derived for 5d SCFTs with holographic duals in Type IIB supergravity.
These theories have relevant deformations that flow to linear quiver gauge theories, 
with the number of nodes large in the large-$N$ limits described by supergravity.
Starting from a suitable formulation of the matrix models resulting from supersymmetric localization of the squashed $S^5$ partition functions, the saddle point equations are solved for generic quivers with $N_f=2N$ at all interior nodes, which includes the $T_N$ theories, and for a sample of theories with $N_f\neq 2N$ nodes including theories with Chern-Simons terms.
The resulting exact expressions for the free energies and conformal central charges are consistent with supergravity predictions and, where available, with previous numerical field theory analyses.
\end{abstract}

\begin{document}

\title{Exact results for 5d SCFTs of long quiver type}

\author{Christoph F.~Uhlemann} 
\email{uhlemann@physics.ucla.edu}

\affiliation{Mani L.\ Bhaumik Institute for Theoretical Physics\\
Department of Physics and Astronomy\\
University of California, Los Angeles, CA 90095, USA}

\maketitle

\tableofcontents

\baselineskip 15pt

\section{Introduction}

Five-dimensional superconformal field theories (SCFTs) play an interesting role in the general understanding of quantum field theory (QFT). On the one hand, they are hard to construct directly using field theory methods, and are typically defined through constructions in string theory and M-theory \cite{Seiberg:1996bd,Intriligator:1997pq, Aharony:1997ju,Kol:1997fv,Aharony:1997bh,Brandhuber:1997ua,Jefferson:2017ahm,Jefferson:2018irk,Xie:2017pfl,Apruzzi:2018nre,Closset:2018bjz,Apruzzi:2019opn}. 
On the other hand, many of them admit relevant deformations that flow to perturbatively non-renormalizable Lagrangian gauge theories, for which the SCFTs provide strongly-coupled UV fixed points. 
In between the maximal six-dimensional theories and theories in lower dimensions, five-dimensional theories allow for an intriguing interplay between string theory constructions, AdS/CFT and field theory methods, and upon compactification they provide many interesting insights into lower-dimensional theories.

This interplay has been very successful for the 5d $USp(N)$ theories, realized by D4/D8/O8 configurations in Type IIA, which for large $N$ have holographic duals in massive Type IIA supergravity \cite{Brandhuber:1999np,Bergman:2012kr}. These theories have been studied extensively e.g.\ in \cite{Jafferis:2012iv,Bergman:2012qh,Assel:2012nf,Bergman:2013koa,Alday:2015lta,Gutperle:2017nwo,Gutperle:2018axv,Bah:2018lyv,Dibitetto:2018iar,Hosseini:2018uzp,Crichigno:2018adf,Hosseini:2018usu,Choi:2019miv,Penin:2019jlf}.
However, many more 5d gauge theories are believed to have strongly-coupled UV fixed points, and indeed many more theories can be engineered in Type IIB string theory using 5-brane webs \cite{Aharony:1997ju,Kol:1997fv,Aharony:1997bh}, 5-brane webs with 7-branes \cite{DeWolfe:1999hj} and further generalizations \cite{Bergman:2015dpa,Hayashi:2018bkd,Hayashi:2018lyv,Hayashi:2019yxj}. 
Type IIB supergravity solutions describing the near-horizon limits of 5-brane webs and 5-brane webs with 7-branes have been constructed in \cite{DHoker:2016ujz,DHoker:2016ysh,DHoker:2017mds} and \cite{DHoker:2017zwj},\footnote{Earlier studies of the BPS equations can be found in \cite{Apruzzi:2014qva,Kim:2015hya,Kim:2016rhs}, and T-duals of the Type IIA solution were discussed in \cite{Lozano:2012au,Lozano:2013oma}.} and result in large classes of explicit holographic dualities for the corresponding 5d SCFTs.
Various aspects of the dualities have been studied in \cite{Gutperle:2017tjo,Kaidi:2017bmd,Gutperle:2018vdd,Gutperle:2018wuk,Bergman:2018hin,Fluder:2018chf,Kaidi:2018zkx,Hong:2018amk,Malek:2018zcz,Lozano:2018pcp,Chaney:2018gjc,Malek:2019ucd,Chen:2019one,Fluder:2019szh}.
In particular, the free energy was found to scale quartically with the parameter controlling the large-$N$ limit in supergravity \cite{Gutperle:2017tjo}, and this scaling was confirmed numerically using field theory methods \cite{Fluder:2018chf,Fluder:2019szh}.

In this paper we study 5d SCFTs that have holographic duals in Type IIB supergravity using field theory methods.
The theories we consider have relevant deformations that are described by linear quiver gauge theories with $SU(N)$ gauge nodes, 
with bifundamental hypermultiplets connecting adjacent nodes.
In addition there may be fundamental hypermultiplets attached to individual nodes and Chern-Simons terms.
The general quiver diagram is shown in eq.~(\ref{eq:quiver}).
We use the gauge theory deformations and supersymmetric localization to analytically compute the free energies of the 5d SCFTs on the round $S^5$ and on squashed spheres, in the large-$N$ limits described by supergravity.

Localization of the (squashed) 5-sphere partition function has been worked out for generic 5d gauge theories in \cite{Kallen:2012va,Kim:2012ava,Lockhart:2012vp,Imamura:2012xg,Imamura:2012bm}. The path integral is reduced to a matrix integral over the Cartan algebra of the gauge group.
At large $N$, instanton contributions are expected to be suppressed, and the zero-instanton part of the partition function is expected to be captured exactly by a saddle point computation.
However, for the gauge theories of interest here, the number of nodes in the quiver diagrams is large in the large-$N$ limits of interest.
Each gauge node contributes a matrix integral, so that the number of matrix integrals itself is large, and for each gauge node one needs to find an a-priori independent saddle point eigenvalue distribution.\footnote{%
For the free energy of quivers arising from orbifolds of the $USp(N)$ theory discussed in \cite{Jafferis:2012iv}, the saddle point eigenvalue distributions are independent of the gauge node, effectively reducing the problem to a single node. For the theories considered here the saddle point configurations depend non-trivially on the quiver node.} 
We set up a suitable language to describe the localized partition functions for long quiver gauge theories, and cast the saddle point equations in a form akin to a 2d electrostatics problem. 
The saddle point equations are solved analytically for generic quivers where all interior nodes have an effective number of flavors equal to twice the number of colors, $N_f=2N$, and for a sample of theories which also have $N_f\neq 2N$ nodes, leading to exact expressions for the squashed sphere free energies.
This includes the theories for which the $S^5$ free energies were computed numerically in \cite{Fluder:2018chf,Fluder:2019szh}, which provided valuable intuition for the analytic computations shown here. 
Following \cite{Chang:2017cdx,Chang:2017mxc}, the conformal central charge characterizing the energy-momentum tensor two-point function is extracted, and we show that a universal relation between conformal central charge and $S^5$ free energy, found numerically for two example theories in \cite{Fluder:2018chf}, holds for all theories of the long quiver type to be discussed below.

The remaining parts are organized as follows. In sec.~\ref{sec:long-quivers} we discuss the squashed $S^5$ free energies for long quiver gauge theories at a general level, derive the saddle point conditions and establish a relation between the conformal central charge and the $S^5$ free energy. 
In sec.~\ref{sec:saddle-point-sec} the saddle point conditions are summarized and solved explicitly for theories with $N_f=2N$ at all interior nodes.
In sec.~\ref{sec:sample-scfts} we discuss the free energies of a sample of 5d gauge SCFTs that have supergravity duals in Type IIB. We conclude with a discussion in sec.~\ref{sec:discussion}.

\section{5d Partition Functions for long quivers}\label{sec:long-quivers}

This section contains a discussion of partition functions on squashed spheres for 5d long quiver gauge theories with holographic duals in Type IIB.
The theories of interest are linear quiver gauge theories with $SU(N)$ gauge nodes, denoted by $(N)$, with bifundamental hypermultiplets between adjacent gauge nodes and optionally additional fundamental hypermultiplets denoted by $[k]$ for $k$ fundamentals. 
5d $SU(N)$ gauge theories with $N>2$ may also have non-trivial Chern-Simons terms, and we denote the Chern-Simons level $c$, if non-zero, by a subscript, e.g.\ $(N)_c$. The form of the quiver gauge theories then is
\begin{align}\label{eq:quiver}
 (&N_1)_{c_1}-(N_2)_{c_2}-\ldots -(N_{L-1})_{c_{L-1}}-(N_L)_{c_L}
 \nonumber\\
 &\hskip 2mm |\hskip 14mm |\hskip 26mm |\hskip 20mm |
 \\
 &[k_1] \hskip 9mm [k_2] \hskip 19mm [k_{L-1}] \hskip 13mm [k_L]
 \nonumber
\end{align}
In the limits of interest here, the length of the quiver in (\ref{eq:quiver}) is large, $L\gg 1$. 
Moreover, with $t=1,\ldots,L$ labeling the quiver node, at least some $N_t$ are large, of order $L$, but not necessarily all of them.
The term ``large $N$'' will be used to refer to this limit with $L\gg 1$.

To implement the large-$L$  limit, it will be convenient to introduce an effectively continuous parameter $z\in[0,1]$ labeling the gauge node,
\begin{align}\label{eq:z-def}
 z&=\frac{t}{L}~.
\end{align}
The left end of the quiver (\ref{eq:quiver}) corresponds to $z=0$, the right end to $z=1$. 
The data $\lbrace N_t,k_t,c_t\rbrace$ characterizing the quiver is then encoded in functions $N(z)$, $k(z)$, $c(z)$ of the effectively continuous variable~$z$, defined by
\begin{align}\label{eq:cont-rep}
 N(z)&= N_{z L}~, 
 & k(z)&= k_{zL}~, &
 c(z)&=c_{zL}~.
\end{align}
In the examples discussed below $N(z)$ will be a continuous, piece-wise linear, concave function.
Fundamental hypermultiplets and Chern-Simons terms only appear at the isolated nodes where $N(z)$ has kinks,
but for a uniform treatment we nevertheless introduce $k(z)$ and $c(z)$, which will be sums of $\delta$-functions, to encode them.

The effective number of flavors at an interior gauge node, including fundamental hypermultiplets and bifundamental hypermultiplets with adjacent gauge nodes, is $N_{f,t}=N_{t+1}+N_{t-1}+k_t$. The continuous version is
\begin{align}\label{eq:Nf-def}
 N_f(z)&=2N(z)+\frac{1}{L^2}\partial_z^2N(z)+k(z)~.
\end{align}
At nodes where $N(z)$ is smooth and $k(z)$ zero, this effective number of flavors is equal to twice the number of colors, $N_f=2N$.
At nodes where $N(z)$ has a kink, $N_f$ may be smaller or equal to twice the number of colors, depending on the number of fundamental hypermultiplets attached to that node.

The general form of the matrix models resulting from supersymmetric localization of the squashed sphere partition function of 5d gauge theories is reviewed in sec.~\ref{sec:5d-part-gen}. 
A suitable formalism for long quiver gauge theories is set up in sec.~\ref{sec:5d-part-long-quiver}. 
The saddle point conditions will be discussed in three parts: 
Along parts of the quiver with no Chern-Simons terms and $N_f=2N$ at each node, they take the form of a partial differential equation, which is derived in sec.~\ref{sec:saddle}. It is supplemented by boundary conditions at $z=0$ and $z=1$ which are discussed in sec.~\ref{sec:bc}. Finally, junction conditions for nodes with $N_f\neq 2N$ are derived in sec.~\ref{sec:junction}.
A universal large-$N$ relation between the conformal central charge and the $S^5$ free energy is derived in sec.~\ref{sec:CT}.

\subsection{5d partition functions}\label{sec:5d-part-gen}

Supersymmetric 5d gauge theories can be formulated on squashed 5-spheres \cite{Imamura:2012xg,Imamura:2012bm}, for which an explicit metric can be written as
\begin{align}\label{eq:squashed-metric}
 ds^2&=\sum_{i=1}^3\left(d\rho_i^2+\rho_i^2 d\theta_i^2\right)-\frac{1}{1+\sum_{i=1}^3\phi_i^2\rho_i^2}\left(\sum_{i=1}^3\phi_i\rho_i^2 d\theta_i\right)^2~,
\end{align}
with real coordinates $\theta_i\in(0,2\pi)$ and $\rho_i\geq 0$, constrained by $\sum_{i=1}^3\rho_i^2=1$. 
The $\phi_i$ are the squashing parameters, and for $\phi_1=\phi_2=\phi_3=0$ the metric reduces to the round $S^5$.
For generic $\phi_i$ the isometry is reduced from $SO(6)$ to $U(1)^3$.
The perturbative part of the squashed $S^5$ partition function has been derived in~\cite{Imamura:2012bm}. For a 5d gauge theory with gauge group $G$ and $N_f$ hypermultiplets in a real representation $R_f \otimes \bar R_{f}$ of $G$, it is given by
\begin{align}\label{eqn:generalpartfunc}
\cZ_{\vec{\omega}}  \ = \ & 
\frac{S_{3}^{\prime} \left( 0\mid \vec{\omega} \right)^{\mathrm{rk} \, G}}{\left| \cW \right| (2\pi)^{\mathrm{rk} \, G}} 
\left[ \prod_{i=1}^{\mathrm{rk} \, G} \int_{-\infty}^{\infty} d \lambda_i \right]e^{-\frac{(2\pi)^{3}}{\omega_1\omega_2\omega_3}\mathfrak{F}(\lambda)}
\times 
\frac{\prod_{\alpha} S_3 \left( -i \alpha(\lambda) \mid \vec{\omega} \right)}{\prod_{f=1}^{N_f} \prod_{\rho_{f}} S_{3} \left( i \rho_{f}(\lambda) + \tfrac{\omega_{\rm tot}}{2} \mid \vec{\omega}\right)} ~.
\end{align}
It depends on the squashing parameters through the periods $\omega_i=1+i\phi_i$, which are collected in $\vec{\omega}=(\omega_1,\omega_2,\omega_3)$, with $\omega_{\rm tot}\equiv\omega_1+\omega_2+\omega_3$.
For the round $S^5$ with $\vec{\omega}=(1,1,1)$ we will also use $S^5$ as a subscript.
The roots of $G$ are denoted by $\alpha$, the flavor hypermultiplets are labeled by $f = 1, \ldots N_f$, and $\rho_f$ are the weights of the corresponding representation $R_{f}\otimes \bar R_{f}$. 
$S_3 (z \mid \vec{\omega})$ is the triple sine function, 
$\cW $ the Weyl group of $G$, and $\mathfrak{F}(\lambda)$ is the classical flat space prepotential. 
At the UV fixed point, where $g_{\rm YM}\rightarrow\infty$, only Chern-Simons terms remain, 
\begin{align}
\mathfrak{F}(\lambda)=\frac{c}{6(2\pi)^2}\tr(\lambda^3)~.
\end{align}
The triple sine function can be represented as
\begin{align}
S_3 \left( z \mid \vec{\omega}\right) \ &= \ \exp \left( 
- \frac{\pi i}{6} B_{3,3} \left( z \mid \vec{\omega} \right) 
- \cI_{3} \left( z \mid \vec{\omega}\right)\right) ~,
\nonumber\\
\cI_{3} \left( z \mid \vec{\omega} \right) \ &= \ \int_{\mathbb{R}+ i 0^{+}} \frac{d x}{x} \frac{e^{zx}}{ \left( e^{\omega_1 x} -1 \right)\left( e^{\omega_2 x} -1 \right)\left( e^{\omega_3 x} -1 \right)} ~,
\nonumber\\
B_{3,3} (z \mid \vec{\omega}) \ &= \ 
\frac{1}{\omega_1\omega_2\omega_3} \bigg[ 
\left(z-\frac{\omega_{\rm tot}}{2}\right)^3-\frac{1}{4}(\omega_1^2+\omega_2^2+\omega_3^2)\left(z-\frac{\omega_{\rm tot}}{2}\right)
\bigg] ~,
\end{align}
where $B_{3,3}$ is a generalized Bernoulli polynomial.
The contour in $\cI_3$ runs over the real axis, avoiding the origin via a semi-circle around $x=0$ going into the positive half-plane.

\subsection{Long linear quiver gauge theories}\label{sec:5d-part-long-quiver}

The zero-instanton partition function (\ref{eqn:generalpartfunc}) for generic quiver gauge theories of the form (\ref{eq:quiver}) can be written conveniently as
\begin{align}\label{eqn:ZMN-gen}
\cZ_{\vec\omega} &= \frac{S_{3}^{\prime} \left( 0\mid \vec{\omega} \right)^{\mathrm{rk} \, G}}{\left| \cW \right| (2\pi)^{\mathrm{rk} \, G}} \int_{-\infty}^{\infty} \left[ \prod_{t=1}^{L} \prod_{i=1}^{N_t-1} d \lambda_{i}^{(t)}  \right]
\exp \left( - \frac{1}{\omega_1\omega_2\omega_3}\cF_{\vec\omega}\right) \,,
\end{align}
with
\begin{align}\label{eq:cF-gen}
 \cF_{\vec{\omega}} \ = \ &
 \sum_{t=1}^L\sum_{\stackrel{\ell,m=1}{\ell\neq m}}^{N_t}F_V\big(\lambda_\ell^{(t)}-\lambda_m^{(t)}\big)
 +\sum_{t=1}^{L-1}\sum_{\ell=1}^{N_t}\sum_{m=1}^{N_{t+1}}F_H\big(\lambda_\ell^{(t)}-\lambda_m^{(t+1)}\big)
  \nonumber\\ &
 +\sum_{t=1}^L k_t\sum_{i=1}^{N_t}F_H\big(\lambda_i^{(t)}\big)
 +\sum_{t=1}^L\frac{\pi}{3}c_t\left(\lambda_\ell^{(t)}\right)^3
 ~,
\end{align}
where
\begin{align}
 F_V(x)&\equiv-\frac{1}{2}\omega_1\omega_2\omega_3\left[\ln S_3\left(ix|\vec{\omega}\right)+\ln S_3\left(-ix|\vec{\omega}\right)\right]~,
\nonumber\\
F_H(x)&\equiv \omega_1\omega_2\omega_3\ln S_3\left(i x+\frac{\omega_{\rm tot}}{2}\mid \vec{\omega}\right)
 ~.
\end{align}
We will need the asymptotic behavior of $F_H$ and $F_V$ for large arguments. The relevant asymptotics of the triple sine function were collected in appendix A of \cite{Chang:2017mxc}. With large real $|x|$,
\begin{align}
 F_V(x)
 & \approx 
 +\frac{\pi}{6}|x|^3
 -\frac{\omega_{\rm tot}^2+\omega_1 \omega_2+\omega_1\omega_3+\omega_2\omega_3}{12}\pi|x|~,
 \nonumber\\
 F_H(x)&\approx-\frac{\pi}{6}|x|^3-\frac{\omega_1^2+\omega_2^2+\omega_3^2}{24}\pi|x|~.
\label{eq:FH-FV-exp}
\end{align}
In the large-$N$ limit instanton constributions are expected to be suppressed, such that the perturbative part captures the large-$N$ behavior, and the saddle point approximation is expected to be exact.
To evaluate the partition functions we introduce normalized eigenvalue densities $\rho_t$ for the $t^{\rm th}$ gauge node, such that
\begin{align}
 \frac{1}{N_t}\sum_{\ell=1}^{N_t} f(\lambda_\ell^{(t)})& \quad\longrightarrow\quad
 \int d\lambda\, \rho_t(\lambda) f(\lambda)~.
\end{align}
In the examples to be discussed below the rank may not be large for all gauge groups, but we can nevertheless introduce the densities for gauge groups of small rank and approximate them by smooth functions.
$\cF_{\vec{\omega}}$ in eq.~(\ref{eq:cF-gen}) then becomes
\begin{align}\label{eq:cF-gen-2}
\cF_{\vec{\omega}} \ = \ &
 \int d\lambda\, d\tilde\lambda\,\left[
 \sum_{t=1}^L N_t^2\rho_t(\lambda)\rho_t(\tilde\lambda)F_V\big(\lambda-\tilde\lambda\big)
 +\sum_{t=1}^{L-1}N_tN_{t+1}\rho_t(\lambda)\rho_{t+1}(\tilde\lambda)F_H\big(\lambda-\tilde\lambda\big)
 \right]
 \nonumber\\ &
 +\int d\lambda \sum_{t=1}^LN_t\rho_t(\lambda)\left[k_tF_H(\lambda)+\frac{\pi}{3}c_t\lambda^3\right]~.
\end{align}
At leading order for large arguments, $F_H$ and $F_V$ are opposite-equal.
This suggests to combine the two terms in the first line to form differences of the eigenvalue distributions at the $t$ and $t+1$ nodes. 
To this end we introduce
\begin{align}\label{eq:F0-def}
 F_0(x)&\equiv \frac{F_V(x)+F_H(x)}{\omega_{\rm tot}^2}~, & 
 \Delta_t(\lambda)&\equiv N_{t+1}\rho_{t+1}(\lambda)-N_t\rho_t(\lambda)~,
\end{align}
where a factor of $\omega_{\rm tot}^{-2}$ has been included in the definition of $F_0$ for later convenience.
Then (\ref{eq:cF-gen-2}) can be written as
\begin{align}\label{eq:cF-gen-3}
\cF_{\vec{\omega}} =\, &
 \int d\lambda\, d\tilde\lambda\,\left[
 \sum_{t=1}^L N_t^2\rho_t(\lambda)\rho_t(\tilde\lambda)\omega_{\rm tot}^2F_0\big(\lambda-\tilde\lambda\big)
 -\frac{1}{2}\sum_{t=1}^{L-1}\Delta_t(\lambda)\Delta_t(\tilde\lambda)F_H\big(\lambda-\tilde\lambda\big)
 \right]
 \nonumber\\ &
  -\frac{1}{2}\sum_{t\in\lbrace 1,L\rbrace}\int\!d\lambda \,d\tilde\lambda\,N_t^2\rho_t(\lambda)\rho_t(\tilde\lambda)
  F_H\big(\lambda-\tilde\lambda\big)
 +\int \! d\lambda \sum_{t=1}^LN_t\rho_t(\lambda)\left[k_tF_H(\lambda)+\frac{\pi}{3}c_t\lambda^3\right].
\end{align}
Note the remainder terms at the first and last node in the second line.
To implement large $L$, a continuous parameter $z\in[0,1]$ as defined in (\ref{eq:z-def}) is introduced to label the gauge nodes, such that the quiver data is encoded in the functions $N(z)$, $k(z)$ and $c(z)$ defined in (\ref{eq:cont-rep}). 
Likewise, the family of eigenvalue densities $\lbrace\rho_t\rbrace$ is replaced by one function of two continuous parameters, 
\begin{align}\label{eq:cont-rep-rho}
 \rho(z,\lambda)&=\rho^{}_{z L}(\lambda)~,
 & 
 z&=\frac{t}{L}~.
\end{align}
With the definitions in (\ref{eq:cont-rep}), (\ref{eq:cont-rep-rho}) the sums over quiver nodes are replaced by integrals,
\begin{align}
 \sum_{t=1}^{L} f_t \ &\rightarrow \ L\int_0^1 dz f(z L)~,
\end{align}
and the finite difference terms in (\ref{eq:cF-gen-3}) turn into derivatives with respect to $z$,
\begin{align}
 \Delta_{zL}(\lambda)
 &\rightarrow \ \frac{1}{L}\partial_z\left[N(z) \rho(z,\lambda)\right]~.
\end{align}
The expression for $\cF_{\vec{\omega}}$ in (\ref{eq:cF-gen-3}) becomes
\begin{align}\label{eq:cF-gen-4}
 \cF_{\vec{\omega}} =\, &
 L\int_0^1 dz \int d\lambda\,d\tilde\lambda\,\Big[N(z)^2\rho(z,\lambda)\rho(z,\tilde\lambda)\omega_{\rm tot}^2F_0(\lambda-\tilde\lambda)
 \nonumber\\ & \hskip 30mm
 -\frac{1}{2L^2}\partial_z\big(N(z)\rho(z,\lambda)\big)\partial_z\big(N(z)\rho(z,\tilde\lambda)\big)F_H(\lambda-\tilde\lambda)\Big]
 \nonumber\\ & 
 -\frac{1}{2}\!\sum_{z\in\lbrace 0,1\rbrace}\int d\lambda\,d\tilde\lambda\,N(z)^2\rho(z,\lambda)\rho(z,\tilde\lambda)F_H(\lambda-\tilde\lambda)
 \nonumber\\ & 
 +L\int_0^1dz\!\int\! d\lambda\, N(z)\rho(z,\lambda)\left[k(z)F_H(\lambda)+\frac{\pi}{3}c(z)\lambda^3\right]~.
\end{align}

To determine the scaling of the eigenvalues in the large-$N$ limits described below (\ref{eq:quiver}), we focus on the first integral in (\ref{eq:cF-gen-4}), assuming that both terms in the square brackets are non-vanishing.
The first term in the square brackets is linear in the eigenvalues and the second term is cubic:
as consequence of the definition in (\ref{eq:F0-def}), the cubic terms in $F_0(\lambda-\tilde \lambda)$ cancel and the explicit expressions for the leading-order behavior for large argument are given by
\begin{align}\label{eq:FH-DeltaF-LO}
F_0(x)&=-\frac{\pi}{8}|x|~,
&
 F_H(x)&=-\frac{\pi }{6}|x|^3~.
\end{align}
Both terms in the square brackets in the first integral in (\ref{eq:cF-gen-4}) are $\mathcal O(N(z)^2)$ without scaling the eigenvalues, so the eigenvalues are not expected to scale non-trivially with $N(z)$.
This leaves the scaling with $L$. 
Assume that the eigenvalues scale as $\lambda = L^\alpha x$, with $x$ of order 1.
Then the first term in the square brackets scales as $L^\alpha$, and the second term as $L^{3\alpha-2}$.
They can combine non-trivially if $\alpha=1$, such that the eigenvalues scale linearly with $L$.
We thus introduce 
\begin{align}\label{eq:lambda-scaling}
 \lambda&=L \omega_{\rm tot} x~, & 
 \hat \rho(z,x)&=L\omega_{\rm tot}\rho(z,L\omega_{\rm tot} x)~,
\end{align}
with $x$ of order one and $\hat\rho$ a normalized density with $\hat\rho(z,x)dx=\rho(z,\lambda)d\lambda$.
The factor $\omega_{\rm tot}$, which is order one, does not affect the scaling but has been included to isolate the dependence on the squashing parameters.
Keeping only the leading terms, we find
\begin{align}
 \cF_{\vec{\omega}}&=\omega_{\rm tot}^3\cF~,
\end{align}
where $\cF$ is independent of the squashing parameters and given by
\begin{align}\label{eq:cF-gen-5}
 \cF \ = \ &
 L^2\int_0^1 dz \int dx\,dy\,\cL
 -\frac{1}{2}L^3\sum_{z\in\lbrace 0,1\rbrace}\int dx\,dy\, N(z)^2\hat\rho(z,x)\hat\rho(z,y)F_H\big(x-y\big)
 \nonumber\\ & 
 +L^4\int_0^1dz\int dx\, N(z)\hat\rho(z,x)\left[k(z)F_H(x)+\frac{\pi}{3}c(z)x^3\right]~,
\end{align}
with
\begin{align}\label{eq:cL}
 \cL&=N(z)^2\hat\rho(z,x)\hat\rho(z,y)F_0(x-y\big)
 -\frac{1}{2}\partial_z \big(N(z)\hat\rho(z,x)\big)\partial_z\big(N(z)\hat\rho(z,y)\big)F_H\big(x-y\big)~,
\end{align}
and $F_0$ and $F_H$ as given in (\ref{eq:FH-DeltaF-LO}).
The requirements for proper normalization of the eigenvalue distributions and the constraint that the eigenvalues sum to zero amount to 
\begin{align}\label{eq:constraints}
 \int dx\,\hat\rho(z,x)&=1~, & \int dx\,x\, \hat\rho(z,x)&=0~.
\end{align}
This calls for a constrained extremization of $\cF$ to determine the saddle points. 
For some of the theories considered below an unconstrained extremization leads to solutions which already satisfy the constraints, such that working with (\ref{eq:cF-gen-5}) is sufficient, but not for all of them. 
The constraints can be implemented by Lagrange multiplier terms.
For the two sets of constraints, each labeled by the continuous parameter $z$, two Lagrange multiplier functions $\mu(z)$ and $\tau(z)$ are introduced, with factors of $L$ and $N(z)$ included in their definition for later convenience. This leads to
\begin{align}\label{eq:cF-gen-6}
 \cF \ = \ &
 L^2\int_0^1 dz \int dx\,dy\,\cL
 -\frac{1}{2}L^3\sum_{z\in\lbrace 0,1\rbrace}\int dx\,dy\, N(z)^2\hat\rho(z,x)\hat\rho(z,y)F_H\big(x-y\big)
 \nonumber\\ & 
 +L^4\int_0^1dz\int dx\, N(z)\hat\rho(z,x)\left[k(z)F_H(x)+\frac{\pi}{3}c(z)x^3\right]
 \nonumber\\
 &+L^4\int_0^1 dz\, N(z)\left[\mu(z)\left(\int dx\, \hat\rho(z,x)-1\right)+\tau(z)\int dx\,x\, \hat\rho(z,x)\right]~.
\end{align}
In summary, the leading large-$N$ version of the partition function $\cZ_{\vec{\omega}}$ in (\ref{eqn:ZMN-gen}) becomes
\begin{align}
 \cZ_{\vec{\omega}}&=
 \int \mathcal D \hat\rho\, \exp\left(-\frac{\omega_{\rm tot}^3}{\omega_1\omega_2\omega_3}\cF\right)~,
\end{align}
with $\cF$ given in (\ref{eq:cF-gen-6}) and $\cL$ in (\ref{eq:cL}), and $\hat\rho$ a distribution on $[0,1]\times\RR$.
The overall constants in (\ref{eqn:ZMN-gen}) are subleading in the large $N$ limits of interest and have been dropped, along with other subleading terms.
The free energy in the saddle point approximation is given by
\begin{align}\label{eq:free-energy-gen}
 F_{\vec{\omega}}\equiv -\ln \cZ_{\vec{\omega}}\approx \frac{\omega_{\rm tot}^3}{\omega_1\omega_2\omega_3}\cF\big\vert_{\hat\rho=\hat\rho_s}~,
\end{align}
with $\hat\rho_s$ denoting the saddle point configuration.
For the round $S^5$ with $\vec{\omega}=(1,1,1)$, the overall factor on the right hand side in (\ref{eq:free-energy-gen}) evaluates to $27$.

\subsection{Saddle point equation}\label{sec:saddle}

To derive the saddle point conditions, the expression for $\cF$ in eq.~(\ref{eq:cF-gen-6}) is first simplified by introducing a rescaled eigenvalue distribution $\varrho$,
\begin{align}\label{eq:varrho-def}
 \varrho(z,x)&\equiv N(z)\hat\rho(z,x)~,
 &
 \int dx\,\varrho(z,x)&=N(z)~.
\end{align}
Then
\begin{align}\label{eq:cF-gen-7}
 \cF = \, &
 L^2\int_0^1 dz \int dx\,dy\, \cL
 -\frac{1}{2}L^3\sum_{z\in\lbrace 0,1\rbrace}\int dx\,dy\, \varrho(z,x)\varrho(z,y)F_H\big(x-y\big)
 \nonumber\\ &
 +L^4\int_0^1dz\int dx\, \varrho(z,x)\left[k(z)F_H(x)+\frac{\pi}{3}c(z)x^3\right]
 \nonumber\\
 &+L^4\int_0^1 dz\,\left[\mu(z)\left(\int dx\, \varrho(z,x)-N(z)\right)+\tau(z)\int dx\,x\, \varrho(z,x)\right]~,
\end{align}
with
\begin{align}
 \cL&=\varrho(z,x)\varrho(z,y) F_0(x-y\big)
 -\frac{1}{2}\partial_z \varrho(z,x)\partial_z\varrho(z,y)F_H\big(x-y\big)~.
\end{align}

The saddle point equation derived in this section is obtained from (\ref{eq:cF-gen-7}) by varying $\varrho$ in a part of the interior of the interval $z\in[0,1]$ along which $\varrho$ is assumed to be smooth. The resulting condition is local in $z$ and reads
\begin{align}\label{eq:rho-eom-0}
 \int dy\left[2\varrho(z,y)F_0(x-y)+\partial_z^2\varrho(z,y)F_H(x-y)\right]
 \nonumber\\ 
 +L^2\left[k(z)F_H(x)+\frac{\pi}{3}c(z)x^3+\mu(z)+x\tau(z)\right]&=0~.
\end{align}
Using the explicit expressions for the leading large-argument behavior of $F_0$ and $F_H$ in (\ref{eq:FH-DeltaF-LO}) and integration by parts in the first term of the integral, along with the required fall-off behavior of $\varrho(z,x)$ for large $x$, this may be rewritten as
\begin{align}\label{eq:rho-eom}
 \int dy\,\Big[\frac{1}{4}\partial_y^2\varrho(z,y)+\partial_z^2\varrho(z,y)+L^2k(z)\delta(y)\Big] F_H(x-y)
 +L^2\Big[\frac{\pi}{3}c(z)x^3+\mu(z)+x\tau(z)\Big]&=0\,.
\end{align}

The saddle point equation (\ref{eq:rho-eom}) is consistent for large $|x|$ along parts of the quiver
without Chern-Simons terms and where the effective number of flavors $N_f(z)$ defined in (\ref{eq:Nf-def}) at each node is equal to twice the number of colors: The leading terms at large $x$ in (\ref{eq:rho-eom}) are cubic, with $F_H(x)\approx -\frac{\pi}{6}|x|^3$ independent of $y$. 
The first term in the integral thus drops out at cubic order, due to the required fall-off behavior of $\varrho$. 
In the second term one may exchange the integration and the derivative $\partial_z^2$, and use the normalization of $\varrho$ to arrive at
\begin{align}\label{eq:rho-eom-large-x-2}
 -\frac{\pi}{6}|x|^3\left[\partial_z^2N(z)+L^2k(z)\right]+\frac{\pi}{3}L^2c(z)x^3&=0~.
\end{align}
This is consistent for large positive and negative $x$ in regions where $c(z)=0$ and $\partial_z^2N(z)=-L^2k(z)$, and the latter is precisely the condition that the effective number of flavors is twice the number of colors, also at nodes where $N(z)$ may have a kink.
In these regions (\ref{eq:rho-eom}) can be imposed for all $x\in\RR$. A local condition can then be derived by acting on (\ref{eq:rho-eom}) with $(\partial_x)^4$, leading to
\begin{align}\label{eq:rho-eom-pde}
 \frac{1}{4}\partial_x^2\varrho(z,x)+\partial_z^2\varrho(z,x)+L^2k(z)\delta(x)&=0~.
\end{align}
Using this condition in (\ref{eq:rho-eom}) shows that, in these regions, the functions $\mu(z)$ and $\tau(z)$ vanish.

\subsection{Boundary conditions}\label{sec:bc}

To derive the boundary conditions at $z=0$ and $z=1$, let $z_b\in\lbrace 0,1\rbrace$.
We assume that $N(z)$ is smooth in a neighborhood of $z_b$, such that (\ref{eq:rho-eom-pde}) holds, 
and that there are $k_b$ fundamental flavors at the boundary node with Chern-Simons level $c_b$.
Since the continuous version of $\delta_{t,t_0}$ is $L^{-1}\delta(z-\frac{t_0}{L})$, $k(z)$ and $c(z)$ become
\begin{align}
 k(z)&=\frac{k_b}{L}\delta(z-z_b)+\ldots~, & c(z)&=\frac{c_b}{L}\delta(z-z_b)+\ldots~,
\end{align}
where the integral over $z$ in (\ref{eq:cF-gen-6}) is understood to include the end points, in the sense that $\delta(z)$ and $\delta(z-1)$ contribute.
Likewise, the Lagrange multiplier functions take the form
\begin{align}
  \mu(z)&=\frac{\mu_b}{L}\delta(z-z_b)+\ldots~, & \tau(z)&=\frac{\tau_b}{L}\delta(z-z_b)+\ldots~.
\end{align}
The variation of $\cF$ in (\ref{eq:cF-gen-6}), with $\delta\hat\rho(z,x)$ non-vanishing only in the aforementioned neighborhood of $z=z_b$, reads
\begin{align}\label{eq:delta-cF-bndy}
\delta \cF=L^2N(z_b)\int dx\,\delta\hat\rho(z_b,x)\Bigg[&
-\int dy \left[n_b\partial_z\left(N(z)\hat\rho(z,y)\right)\big\vert_{z=z_b}+L N(z_b)\hat\rho(z_b,y)\right]F_H(x-y)
\nonumber\\ &
+L \left(k_b F_H(x)+\frac{\pi}{3}c_b x^3+\mu_b+\tau_b x\right)
\Bigg].
\end{align}
The first term in the inner square brackets in (\ref{eq:delta-cF-bndy}) results from integration by parts in $\cL$, 
where $n_b$ is the outward-pointing unit vector normal to the boundary, $n_b=1$ for $z_b=1$ and $n_b=-1$ for $z_b=0$.
The second term is due to the explicit boundary terms in (\ref{eq:cF-gen-6}).

The argument for the boundary conditions depends on whether $N(z)$ is non-zero or zero as $z\rightarrow z_b$:
If $N(z_b)>0$, the first term in the inner square bracket is subleading with respect to the second one, and the boundary conditions are determined from the vanishing of the leading terms.
The case that will be encountered below is 
\begin{align}\label{eq:rho-bc-gen-1}
 N(z_b)&>0~, & k_b&=N(z_b)+\mathcal O(1)~, & c_b&=0~.
\end{align}
That is, the rank of the boundary gauge node is large, with no Chern-Simons term, but with a number of fundamental hypermultiplets which is large as well and differs from the rank of the gauge group $N(z_b)$ only by an order one number. 
That means the effective number of flavors at the boundary node, including the bifundamentals with the adjacent gauge node, to leading order is equal to twice the number of colors.
The leading-order part of the condition $\delta\cF=0$ with (\ref{eq:delta-cF-bndy}) and only $N(z_b)$ and $k_b$ non-vanishing then reduces to 
\begin{align}
-\int dy\, L N(z_b)\hat\rho(z_b,y)F_H(x-y)+L k_b F_H(x)&=0~.
\end{align}
For $k_b=N(z_b)$ this is solved by
\begin{align}\label{eq:rho-bc-gen-2}
 \hat\rho(z_b,x)&=\delta(x)~.
\end{align}

If $N(z)$ vanishes as $z\rightarrow z_b$, the boundary terms in $\delta \cF$ in (\ref{eq:delta-cF-bndy}) vanish, and there are no constraints from extremality of $\cF$. This case corresponds to a quiver tail along which the rank of the gauge groups decreases from order $L$ to order one. In this case, the relation between $\hat\rho$ and $\varrho$ in (\ref{eq:varrho-def}) becomes singular. 
The requirement that the eigenvalue distributions $\hat\rho$ be well behaved at $z=z_b$ is therefore non-trivial, and regular eigenvalue distributions at $z=z_b$ can be obtained if 
\begin{align}\label{eq:rho-bc-gen-3}
  \lim_{z\rightarrow z_b}N(z)\hat\rho(z,x)=\varrho(z_b,x)&=0~.
\end{align}
This condition can also be deduced from the normalization condition in (\ref{eq:varrho-def}).
Both scenarios, (\ref{eq:rho-bc-gen-2}) and (\ref{eq:rho-bc-gen-3}), are summarized by
\begin{align}\label{eq:rho-bc-gen}
 \varrho(z_b,x)&=N(z_b)\delta(x)~.
\end{align}
With these boundary conditions the explicit boundary terms in the first line of (\ref{eq:cF-gen-6}) 
and contributions from fundamental flavors at the boundary nodes drop out.

\subsection{Junction conditions}\label{sec:junction}

As discussed in sec.~\ref{sec:saddle}, the saddle point equation (\ref{eq:rho-eom}) is consistent for large positive and large negative $x$ only along nodes with no Chern-Simons terms and effective number of flavors equal to twice the number of colors.
At (isolated) nodes which are not of this type, two solutions to the local condition (\ref{eq:rho-eom-pde}) are joined, and we now derive the junction conditions.

Let $z_t\in(0,1)$ label an interior node where $N(z)$ has a kink, with $k_t$ fundamental hypermultiplets and a Chern-Simons term with level $c_t$,
\begin{align}
 k(z)&=\frac{k_t}{L}\delta(z-z_t)+\ldots~, & c(z)&=\frac{c_t}{L}\delta(z-z_t)+\ldots~,
 \nonumber\\
 \mu(z)&=\frac{\mu_t}{L}\delta(z-z_t)+\ldots~, & \tau(z)&=\frac{\tau_t}{L}\delta(z-z_t)+\ldots~.
\end{align}
The junction condition for the local solutions in the regions $z<z_t$ and $z>z_t$ are derived from extremality of $\cF$ in (\ref{eq:cF-gen-6}),
by using eq.~(\ref{eq:rho-eom}) in the regions $z<z_t$ and $z>z_t$ separately.
For a variation $\delta\hat\rho(z,x)$ which is non-vanishing only in a region around $z_t$ in which $z_t$ is the only kink, the variation of $\cF$ reads
\begin{align}\label{eq:delta-cF-junction}
 \delta \cF&=L^2\int_0^1dz \int dx\,dy\,\delta\cL+L^3N(z_t)\int dx\,\delta\hat\rho(z_t,x)\left[k_tF_H(x)+\frac{\pi}{3}c_tx^3+\mu_t+\tau_t x\right]
~.
\end{align}
Using (\ref{eq:rho-eom-pde}) with integration by parts and the required fall-off behavior of $\hat\rho(z,y)$ for large $y$ in the first term, the condition $\delta\cF=0$ leads to
\begin{align}\label{eq:junction-gen}
 \int dy\,\left[\partial_z \varrho(z,y)\right]_{z=z_t-\epsilon}^{z=z_t+\epsilon}F_H(x-y)
 +Lk_t F_H(x)+\frac{\pi}{3}Lc_t x^3+L\mu_t+L\tau_t x&=0~,
\end{align}
which has to be imposed for all $x$ for which $\hat\rho$ and thus $\delta\hat\rho$ are allowed to be non-vanishing.
For large $|x|$, (\ref{eq:junction-gen}) leads to the condition found in (\ref{eq:rho-eom-large-x-2}) before.
Depending on the value of the Chern-Simons level, the support of $\varrho$ thus has to be bounded from below, from above, or both, 
by Dirichlet boundary conditions as follows
\begin{align}\label{eq:junction-constr}
 \varrho(z_t,x)&=0 \quad \forall x\lessgtr x_0 & c_t&=\pm \frac{1}{2}\left(\frac{1}{L}\left[\partial_z N(z)\right]_{z=z_t-\epsilon}^{z=z_t+\epsilon}+k_t\right)~,
 \nonumber\\
 \varrho(z_t,x)&=0 \quad \forall x\notin (x_0,x_1) & |c_t|&<\frac{1}{2}\left| \frac{1}{L}\left[\partial_z N(z)\right]_{z=z_t-\epsilon}^{z=z_t+\epsilon}+k_t\right|~.
\end{align}
Acting with $(\partial_x)^4$ on (\ref{eq:junction-gen}) shows that $\partial_z\varrho(z,x)$ is continuous for $x\neq 0$ in the intervals where $\varrho$ is not constrained by (\ref{eq:junction-constr}), with a source at $x=0$ if fundamental flavors are present.
The bounds $x_0$ and $x_1$ in (\ref{eq:junction-constr}) will be determined by the constraints (\ref{eq:constraints}) and the junction condition (\ref{eq:junction-gen}).

\subsection{Conformal central charge \texorpdfstring{$C_T$}{CT}}\label{sec:CT}

We now derive a general relation between the S$^5$ free energy and the conformal central charge $C_T$ for the type of theories discussed in the previous section.
It follows from the general form of the free energy in (\ref{eq:free-energy-gen}) and does not need an explicit solution to the saddle point conditions.
Following \cite{Chang:2017cdx,Chang:2017mxc} (see also \cite{Bobev:2017asb}), the conformal central charge $C_T$ can be computed from the squashed sphere free energy. 
Namely, with $\omega_i=1+a_i$ it can be obtained from an expansion for small $a_i$ via
\begin{align}\label{FS5-CT-gen}
 F_{\left(1+a_1,1+a_2,1+a_3\right)} &= F_{\rm S^{5}} - \frac{\pi^{2} C_T}{1920} \Bigg( \sum_{i=1}^{3} a_i^{2} - \sum_{i<j} a_i a_j \Bigg) + \mathcal O(a_i^{3}) ~.
\end{align}
For the theories discussed above, the squashing parameters enter the large-$N$ free energy only through the overall factor in (\ref{eq:free-energy-gen}), since $\cF$ and consequently also the saddle point equations are independent of $\vec{\omega}$.
The large-$N$ free energy on a generic squashed sphere is therefore related to the free energy on the round $S^5$ by
\begin{align}\label{eq:FSomega-FS5-0}
 F_{\vec{\omega}}&=\frac{\omega_{\rm tot}^3}{27\omega_1\omega_2\omega_3} F_{S^5}~.
\end{align}

Setting $\omega_i=1+a_i$ and expanding (\ref{eq:FSomega-FS5-0}) for small $a_i$ yields
\begin{align}\label{eq:FSomega-FS5}
 F_{\left(1+a_1,1+a_2,1+a_3\right)} &= F_{\rm S^{5}} + \frac{1}{3} \Bigg( \sum_{i=1}^{3} a_i^{2} - \sum_{i<j} a_i a_j \Bigg)F_{\rm S^{5}} + \mathcal O(a_i^{3}) ~.
\end{align}
Comparing (\ref{eq:FSomega-FS5}) and (\ref{FS5-CT-gen}) shows
\begin{align}
C_T&=-\frac{640}{\pi^2}F_{S^5}~.
\end{align}
This is the large-$N$ relation found in the numerical field theory results of \cite{Fluder:2018chf} and derived there in supergravity.

\section{Saddle points from 2d electrostatics}\label{sec:saddle-point-sec}

The combination of the local saddle point equation (\ref{eq:rho-eom-pde}) with the boundary and junction conditions derived in the previous section poses a problem akin to 2d electrostatics. The general problem is summarized in the following. In sec.~\ref{sec:Nf2NC-sol} we solve it explicitly for quiver gauge theories with $N_f=2N$ at all interior gauge nodes. A number of the theories that were considered explicitly so far, e.g.\ in \cite{Bergman:2018hin,Fluder:2018chf}, are of this type, including the $T_N$ and $+_{N,M}$ theories. 
Theories with $N_f\neq 2N$ interior nodes are also included in sec.~\ref{sec:sample-scfts}, in the form of the $Y_N$ and $\pslash_N$ theories, and the corresponding saddle points will be discussed there.

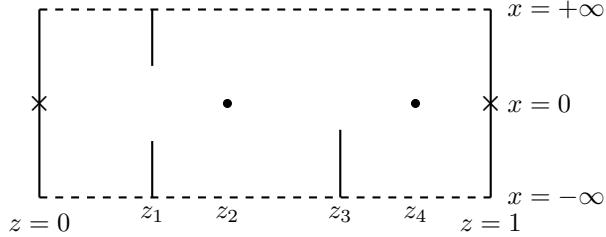
\begin{figure}
\centering 
\begin{tikzpicture}
 \draw[thick] (0,0) -- (0,2.5);
 \draw[thick] (6,0) -- (6,2.5);
 \draw[thick,dashed] (0,0) -- (6,0);
 \draw[thick,dashed] (0,2.5) -- (6,2.5);

 \node [anchor=west] at (6.1,0) {\small $x=-\infty$};
 \node [anchor=west] at (6.1,2.5) {\small $x=+\infty$};
 \node [anchor=west] at (6.1,1.25) {\small $x=0$};
 
 \node [anchor=north] at (0,-0.1) {\small $z=0$};
 \node [anchor=north] at (6,-0.1) {\small $z=1$};
 
 \draw[thick] (1.5,0) -- (1.5,0.75);
 \draw[thick] (1.5,2.5) -- (1.5,1.75);
 \node [anchor=north] at (1.5,0) {\small $z_1$};
 
 \draw [thick] (4,0) -- (4,0.9);
 \node [anchor=north] at (4,0) {\small $z_3$};

 \draw[fill=black] (2.5,1.25) circle (1.5pt);
 \node [anchor=north] at (2.5,0) {\small $z_2$}; 
 
 \draw[fill=black] (5,1.25) circle (1.5pt);
 \node [anchor=north] at (5,0) {\small $z_4$};
 
 \node at (0,1.25) {{\boldmath $\times$}};
 \node at (6,1.25) {{\boldmath $\times$}};
\end{tikzpicture}
\caption{Schematic form of the electrostatic problem associated with a generic quiver.
At $z=0$ and $z=1$ Dirichlet boundary conditions are imposed, which are vanishing aside from $\delta$-functions at the marked points.
The black dots at $x=0$ represent point charges due to fundamental flavors at interior nodes.
The solid lines at $z=z_1$ and $z=z_3$ represent perfectly conducting plates at interior nodes with $N_f\neq 2N$, with total charge given by the discontinuity in $\partial_z N(z)$.
The node at $z=z_3$ has maximal Chern-Simons level, so that the support of $\varrho$ is bounded only on one side; 
the Chern-Simons level at the $z=z_1$ node is smaller.
\label{fig:electrostatics}
}
\end{figure}

As discussed in sec.~\ref{sec:saddle}, along parts of the quiver where $N_f=2N$ with no Chern-Simons terms, 
the support of $\varrho(z,x)$ is unrestricted and it satisfies
\begin{align}\label{eq:rho-eom-pde-rep}
 \frac{1}{4}\partial_x^2\varrho(z,x)+\partial_z^2\varrho(z,x)+L^2k(z)\delta(x)&=0~.
\end{align}
This is a Poisson equation and $\varrho$ can thus be interpreted as an electrostatics potential on the strip $[0,1]\times\RR$, which has to be non-negative.
Fundamental flavor fields account for the source terms given by $k(z)\delta(x)$.
For the theories considered below, there are fundamental flavors only at a finite number of nodes, such that $k(z)$ takes the form
\begin{align}\label{eq:k-z-disc}
 k(z)&=\sum_{t=1}^{L} \frac{k_t}{L}\delta\left(z-z_t\right)~, & z_t&\equiv \frac{t}{L}~,
\end{align}
with only a finite number of terms in the sum non-vanishing. 
The boundary conditions at $z=0$ and $z=1$ as derived in sec.~\ref{sec:bc} are
\begin{align}\label{eq:rho-bc-gen-rep}
 \varrho(0,x)&=N(0)\delta(x)~, & \varrho(1,x)&=N(1)\delta(x)~.
\end{align}
The number of fundamental flavors at the boundary nodes was assumed to be of the same order as the number of colors, which will be the case in all examples considered below.
We thus need an electrostatics potential between two infinite plates with prescribed Dirichlet boundary conditions.
For normalizability of the eigenvalue distributions, $\varrho$ also needs to vanish for $x\rightarrow \pm\infty$.
At interior nodes with $N_f<2N$ there may be Chern-Simons terms, and as discussed in sec.~\ref{sec:junction} the following constraints have to be imposed, depending on the value of the Chern-Simons level $c_t$,
\begin{align}\label{eq:junction-constr-rep}
 \varrho(z_t,x)&=0 \quad \forall x\lessgtr x_0 & c_t&=\pm \frac{1}{2}\left(\frac{1}{L}\left[\partial_z N(z)\right]_{z=z_t-\epsilon}^{z=z_t+\epsilon}+k_t\right)~,
 \nonumber\\
 \varrho(z_t,x)&=0 \quad \forall x\notin (x_0,x_1) & |c_t|&<\frac{1}{2}\left| \frac{1}{L}\left[\partial_z N(z)\right]_{z=z_t-\epsilon}^{z=z_t+\epsilon}+k_t\right|~.
\end{align}
The end points $x_0$ and $x_1$ are determined from the junction condition
\begin{align}\label{eq:junction-gen-rep}
 \int dy\,\left[\partial_z\varrho(z,y)\right]_{z=z_t-\epsilon}^{z=z_t+\epsilon}F_H(x-y)
 +Lk_t F_H(x)+\frac{\pi}{3}Lc_t x^3+L\mu_t+L\tau_t x&=0~,
\end{align}
and from the normalization and  $SU(N)$ constraints in (\ref{eq:constraints}).
Both cases in (\ref{eq:junction-constr-rep}) are compatible with the bound $2|c|\leq 2N-N_f$ of \cite{Intriligator:1997pq}. 
In the electrostatics analogy the conditions in (\ref{eq:junction-constr-rep}) amount to the insertion of semi-infinite, perfectly conducting plates parallel to the plates at $z=0$ and $z=1$.
The charge at generic $z=z_t$ is determined from Gau\ss's law
\begin{align}
 Q_t&=\int dx\,\left[\partial_z\varrho(z,x)\right]_{z=z_t-\epsilon}^{z=z_t+\epsilon}=\left[\partial_z N(z)\right]_{z_t-\epsilon}^{z_t+\epsilon}~,
\end{align}
where the second equality follows from the normalization of $\varrho$.
At nodes where $N(z)$ has a kink, the charge may be provided entirely by fundamental fields, leading to the case $N_f=2N$, 
or at least in part by the conducting plates restricting the support of $\varrho$ at $z=z_t$.
A schematic representation of the electrostatics problem is shown in fig.~\ref{fig:electrostatics}.

\subsection{General solution for \texorpdfstring{$N_f=2N$}{Nf=2N} quivers}\label{sec:Nf2NC-sol}

If all interior nodes have $N_f=2N$, $\varrho$ satisfies the saddle point equation (\ref{eq:rho-eom-pde-rep}) on the entire strip, with boundary conditions given by (\ref{eq:rho-bc-gen-rep}).
With $k(z)$ in (\ref{eq:k-z-disc}) the equation satisfied in the interior of the strip becomes
\begin{align}\label{eq:rho-eom-nf2nc}
 \frac{1}{4}\partial_x^2\varrho(z,x)+\partial_z^2\varrho(z,x)+\sum_{t=2}^{L-1}L k_t\delta\left(z-z_t\right)\delta(x)&=0~.
\end{align}
The flavors at the first and last node of the quiver are crucial in the discussion of boundary conditions and are reflected in (\ref{eq:rho-bc-gen-rep}); they do not play a role for the equation in the interior of the strip.
We solve this equation by mapping the strip to the upper half plane via
\begin{align}
 u&=e^{2\pi x+i\pi z}~.
\end{align}
The boundaries at $z=0$ and $z=1$ are mapped to the positive and negative real line, respectively. The points at the origin and at infinity correspond to large negative and large positive $x$, respectively.
The saddle point equation (\ref{eq:rho-eom-nf2nc}) becomes 
\begin{align}\label{eq:varrho-eom}
 \partial_u\partial_{\bar u}\varrho+\frac{1}{2}L\sum_{t=2}^{L-1} k_t \delta(u,u_t)&=0~, & u_t&=e^{i\pi z_t}~.
\end{align}
That is, the flavors contribute as point charges on the unit circle.
The boundary conditions at $z=0$ and $z=1$ become Dirichlet boundary conditions on the real line,
\begin{align}\label{eq:varrho-bc}
 \varrho\big\vert_{u\in\RR} &= 2\pi N(0)\delta(u,-1)+2\pi N(1)\delta(u,1)~.
\end{align}

Since the problem is linear, a solution can be found by first constructing the solution to (\ref{eq:varrho-eom}) with vanishing Dirichlet boundary condition, and then superimposing a harmonic function to implement (\ref{eq:varrho-bc}). 
The Green's function on the upper half plane with vanishing Dirichlet boundary condition and $\partial_u\partial_{\bar u}G(u,v)=\delta(u,v)$ is given by
\begin{align}
 G(u,v)&=\frac{1}{\pi}\ln\left\vert\frac{u-v}{u-\bar v}\right|^2~.
\end{align}
The solution to (\ref{eq:varrho-eom}) with vanishing Dirichlet boundary condition then is
\begin{align}
 \varrho_0(u)&=\int d^2v\, G(u,v)\left[-\frac{1}{2}L\sum_{t=2}^{L-1} k_t \delta(u,u_t)\right]=-\frac{1}{2}L\sum_{t=2}^{L-1} k_t G\left(u,e^{i\pi z_t}\right)~.
\end{align}
The complete solution for $\varrho$, with an added harmonic function to satisfy (\ref{eq:varrho-bc}), reads
\begin{align}
 \varrho_s(u)&=
 iN(0)\left[\frac{1}{u+1}-\frac{1}{\bar u+1}\right]
 +iN(1)\left[\frac{1}{u-1}-\frac{1}{\bar u-1}\right]
 -\frac{L}{2\pi}\sum_{t=2}^{L-1} k_t \ln\left|\frac{u-u_t}{u-\bar u_t}\right|^2~.
\end{align}
Transforming back to the strip leads to
\begin{align}\label{eq:varrho-s}
 \varrho_s(z,x) = \,&
 \frac{N(0)\sin(\pi z)}{\cosh(2\pi x)-\cos(\pi z)}+\frac{N(1)\sin(\pi z)}{\cosh(2\pi x)+\cos(\pi z)}
 \nonumber\\&
 -\frac{L}{2\pi}\sum_{t=2}^{L-1} k_t
 \ln \left(\frac{\cosh (2 \pi  x)-\cos\left(\pi(z-z_t)\right)}{\cosh (2 \pi  x)-\cos\left(\pi(z+z_t)\right)}\right)
 ~, & z_t&=\frac{t}{L}~.
\end{align}
The actual eigenvalue distributions are obtained via (\ref{eq:varrho-def}),
\begin{align}\label{eq:hat-rho-varrho-s}
 \hat\rho_s(z,x)&=\frac{\varrho_s(z,x)}{N(z)}~.
\end{align}
Since (\ref{eq:varrho-s}) is symmetric under $x\rightarrow -x$, the $SU(N)$ constraint, requiring that the eigenvalues sum to zero, is satisfied.
The norm evaluates to
\begin{align}
 \int dx\, \varrho_s(z,x)&=(1-z)N(0)+zN(1)-\frac{L}{2}\sum_{t=2}^{L-1} k_t\left(|z-z_t|+ 2z_tz -z-z_t\right)~.
\end{align}
This is precisely $N(z)$, as can be seen from the fact that both are piece-wise linear, agree on the boundary values and have identical second derivatives, such that $\hat\rho_s$ is properly normalized.

To obtain the free energies, $\cF$ in (\ref{eq:cF-gen-6}) is evaluated on the saddle point configuration (\ref{eq:hat-rho-varrho-s}) with (\ref{eq:varrho-s}). 
The details are given in app.~\ref{sec:cF-details}, the result is
\begin{align}\label{eq:cF-gen-Nf2NC}
 \cF\big\vert_{\hat\rho=\hat\rho_s} = \, &
 -\frac{L^2}{16\pi^2}\left[2N(0)^2+2N(1)^2+3N(0)N(1)\right]\zeta(3)
 \nonumber\\ &
 -\frac{L^3}{4\pi^3}\sum_{t=2}^{L-1}k_t 
 \left[N(0) D_4\big(e^{i \pi  z_t}\big)+N(1) D_4\big(e^{i \pi (1- z_t)}\big)\right]
 \nonumber\\ &
 +\frac{L^4}{16 \pi^4}\sum_{t=2}^{L-1}\sum_{s=2}^{L-1}k_tk_s\left[D_5\big(e^{i\pi(z_s+z_t)}\big)-D_5\big(e^{i\pi(z_s-z_t)}\big)\right]~,
\end{align}
with the Riemann $\zeta$-function $\zeta(s)={\rm Li}_s(1)$ and the single-valued polylogarithms\footnote{The functions $D_n$ agree, for example, with Zagier's single-valued polylogarithms \cite{Zagier1990} evaluated on a phase.}
\begin{align}\label{eq:D-n-def}
 D_n(e^{i\alpha})&=\begin{cases}
                    \Im\left({\rm Li}_n(e^{i\alpha})\right) & \text{for $n$ even}\\
                    \Re\left({\rm Li}_n(e^{i\alpha})\right) & \text{for $n$ odd}
                \end{cases}~.
\end{align}
The free energy is given by (\ref{eq:free-energy-gen}) with $\cF\vert_{\hat\rho=\hat\rho_s}$ in (\ref{eq:cF-gen-Nf2NC}).

\section{5d SCFTs with gravity duals in Type IIB}\label{sec:sample-scfts}

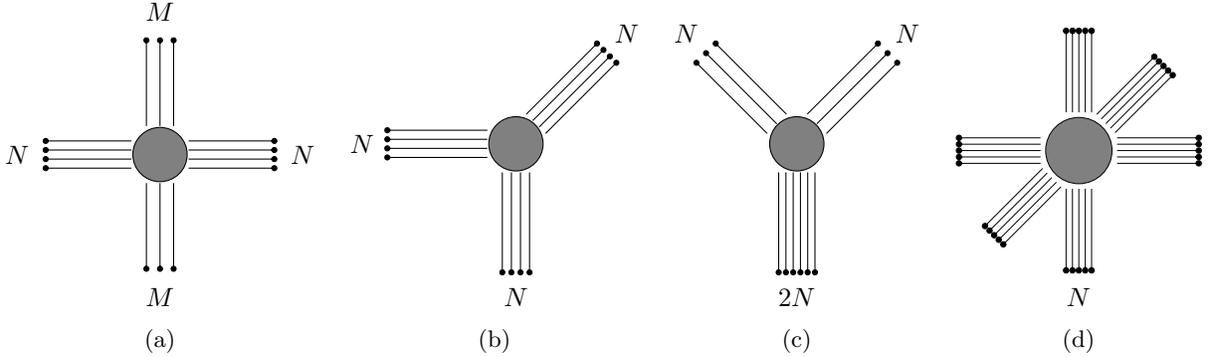
\begin{figure}
\subfigure[][]{\label{fig:plus}
 \begin{tikzpicture}[scale=0.95]
    \foreach \i in {-3/2,-1/2,1/2,3/2}{
      \draw (-0.4,0.127*\i) -- (-1.6,0.127*\i) [fill=black] circle (1pt);
      \draw (0.4,0.127*\i) -- (1.6,0.127*\i) [fill=black] circle (1pt);
    }
    \foreach \i in {-3/2,0,3/2}{
      \draw (0.127*\i,-0.4) -- (0.127*\i,-1.6) [fill=black] circle (1pt);
      \draw (0.127*\i,0.4) -- (0.127*\i,1.6) [fill=black] circle (1pt);
    }
    \draw[fill=gray] (0,0) circle (0.38);
    \node at (0,-2.0) {\small $M$};
    \node at (2.0,0) {\small $N$};
    \node at (0,2.0) {\small $M$};
    \node at (-2.0,0) {\small $N$};
\end{tikzpicture}
}\hskip 0mm
\subfigure[][]{\label{fig:TN}
 \begin{tikzpicture}[scale=0.95]
    \foreach \i in {-3/2,-1/2,1/2,3/2}{
      \draw (-0.4,0.127*\i) -- (-1.8,0.127*\i) [fill=black] circle (1pt);
      \draw (0.127*\i,-0.4) -- (0.127*\i,-1.8) [fill=black] circle (1pt);
      \draw (0.28+0.09*\i,0.28-0.09*\i) -- (1.27+0.09*\i,1.27-0.09*\i) [fill=black] circle (1pt) ;
    }
    \draw[fill=gray] (0,0) circle (0.38);    
    \node at (0,-2.15) {\small $N$};
    \node at (-2.15,0) {\small $N$};
    \node at (1.55,1.55) {\small $N$};
\end{tikzpicture}
}\hskip 0mm
\subfigure[][]{\label{fig:YN}
 \begin{tikzpicture}[scale=0.95]
    \foreach \i in {-2,-6/5,-2/5,2/5,6/5,2}{
      \draw (0.127*\i,-0.4) -- (0.127*\i,-1.8) [fill=black] circle (1pt);
    }\foreach \i in {-3/2,0,3/2}{
      \draw (0.28+0.09*\i,0.28-0.09*\i) -- (1.27+0.09*\i,1.27-0.09*\i) [fill=black] circle (1pt) ;
      \draw (-0.28-0.09*\i,0.28-0.09*\i) -- (-1.27-0.09*\i,1.27-0.09*\i) [fill=black] circle (1pt) ;
    }
    \draw[fill=gray] (0,0) circle (0.38);    
    \node at (0,-2.15) {\small $2N$};
    \node at (1.55,1.55) {\small $N$};
    \node at (-1.55,1.55) {\small $N$};
\end{tikzpicture}
}\hskip 0mm
\subfigure[][]{\label{fig:pslash}
\begin{tikzpicture}
 \node at (0,0) {\includegraphics[width=0.2\linewidth,clip,trim={0 0.3cm 0 0}]{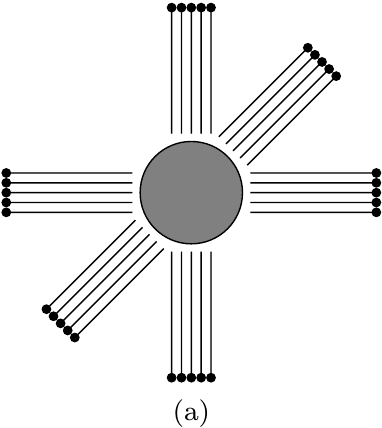} };
 \node at (0,-1.9) {\small $N$};
\end{tikzpicture}
}
 \caption{5-brane junctions for the $+_{N,M}$, $T_N$, $Y_N$ and $\pslash_N$ theories from left to right.
 $(p,q)$ 5-branes are represented by straight lines at angles determined by the $p,q$ charges, the filled black dots represent corresponding $[p,q]$ 7-branes. The 5d SCFTs are realized by intersections at a point; the external 5-branes have been resolved slightly to visually represent the involved branes.\label{fig:unconstrained}}
\end{figure}

In this section the general results of the previous sections are used to compute squashed $S^5$ free energies for a sample of theories with gravity duals in Type IIB.
The 5d SCFTs are engineered in Type IIB string theory by $(p,q)$ 5-brane junctions.
We will consider two classes of theories, distinguished from the string theory perspective by whether or not $s$-rule constraints for multiple 5-branes ending on one 7-brane play a crucial role.
The theories of the first class are shown in fig.~\ref{fig:unconstrained}, those of the second class in fig.~\ref{fig:constrained}.
They all have relevant deformations that flow to long quiver gauge theories of the form (\ref{eq:quiver}) in the IR.
From the field theory perspective they are distinguished by whether or not there are fundamental hypermultiplets at interior gauge nodes.
The $T_N$, $+_{N,M}$, $T_{2K,K,2}$, $T_{N,K,j}$ and $+_{N,M,k}$ theories have $N_f=2N$ at all interior gauge nodes, such that the results follow straightforwardly from sec.~\ref{sec:Nf2NC-sol}.
The $Y_N$ and $\pslash_N$ theories have interior nodes with $N_f\neq 2N$, and the $Y_N$ theory also has a non-trivial Chern-Simons term.

\subsection{\texorpdfstring{$+_{N,M}$}{+[N,M]} theory}

In Type IIB string theory the $+_{N,M}$ theory is defined on the intersection of $N$ D5 and $M$ NS5 branes, fig.~\ref{fig:plus}, and was discussed already in \cite{Aharony:1997bh}.
In field theory it can be defined as the UV fixed point of the linear quiver gauge theory
\begin{align}\label{eq:D5NS5-quiver}
[N]-(N)-\ldots -(N)-[N] ~,
\end{align}
with a total of $M-1$ $SU(N)$ nodes and Chern-Simons levels zero for all gauge nodes.
For $N=M=2$ this is the $E_5$ theory of \cite{Seiberg:1996bd}.
The S-dual gauge theory deformation of the $+_{N,M}$ theory leads to a quiver of the same form, but with $N$ and $M$ exchanged.
The limit described by supergravity corresponds to $N,M\gg 1$ with $N/M$ fixed.
For this gauge theory all gauge groups have large rank in the supergravity limit.
The sphere partition function has been obtained numerically from localization in \cite{Fluder:2018chf} and matched to an analytic supergravity prediction obtained in \cite{Gutperle:2017tjo}.
The matrix model for the quiver (\ref{eq:D5NS5-quiver}) is defined by (\ref{eq:cF-gen-3}) with $N_t=N$ for all $t$ and $L=M-1$.
The only non-vanishing $k_t$ are $k_1=k_{M-1}=N$. 
The continuous versions appearing in (\ref{eq:cF-gen-6}) are
\begin{align}\label{eq:plus-NM-data}
 N(z)&=N~, &  k(z)&=\frac{N}{L}\left(\delta(z)+\delta(z-1)\right)~, & c(z)&=0~.
\end{align}
With a slight abuse of notation we denote by $N(z)$ the continuous function describing the quiver and by $N$ the integer number of D5 branes.
Since $N_f=2N$ at all nodes, the results of sec.~\ref{sec:Nf2NC-sol} apply.
The Lagrange multiplier functions $\mu(z)$ and $\tau(z)$ vanish and do not need to be included in order to find consistent results.
With this data $\cF$ in eq.~(\ref{eq:cF-gen-6}) for the $+_{N,M}$ theory becomes
\begin{align}\label{eqn:cFMN-6}
\cF_{+_{N,M}} &= 
M^2\!\int dz\,dx\,dy\,\cL
+N^2M^3\!\!\sum_{z\in\lbrace 0,1\rbrace}\int dx\, \hat\rho(z,x)\left[F_H(x)-\frac{1}{2}\int dy\, \hat\rho(z,y)F_H(x-y)\right].
\end{align}
The saddle point eigenvalue distribution is given by (\ref{eq:varrho-s}) with the data in (\ref{eq:plus-NM-data}), which is
\begin{align}\label{eq:rho-sol-plus}
 \hat\rho_s(z,x)&=\frac{4\sin (\pi  z) \cosh \left(2 \pi x\right) }{\cosh \left(4 \pi x\right)-\cos (2 \pi  z)}~.
\end{align}
At the center node, $z=\frac{1}{2}$, the eigenvalues are the largest; as the boundaries $z=0$ and $z=1$ are approached, the eigenvalues become concentrated at zero.
The squashed $S^5$ free energy is obtained from (\ref{eq:free-energy-gen}) with the general form of $\cF\vert_{\hat\rho=\hat\rho_s}$ in (\ref{eq:cF-gen-Nf2NC}) and the quiver data in (\ref{eq:plus-NM-data}), which yields
\begin{align}\label{eqn:cFMN-eval-4}
F_{\vec{\omega}}^{\#_{M,N}} &=-\frac{7}{16\pi^2}\frac{\omega_{\rm tot}^3}{\omega_1\omega_2\omega_3}\zeta(3)N^2M^2~.
\end{align}
For the round sphere with $\vec{\omega}=(1,1,1)$, (\ref{eqn:cFMN-eval-4}) agrees with the supergravity result of \cite{Gutperle:2017tjo} and matches the field theory numerics of \cite{Fluder:2018chf}.

One may compare to the results of \cite{Jafferis:2012iv}, where the large-$N$ free energy was computed for the 5d $USp(N)$ theories and their orbifolds introduced in \cite{Bergman:2012kr}.
For example, the quiver obtained from a $\ZZ_{2k}$ orbifold shown in fig.~1(c) of \cite{Jafferis:2012iv} involves the same gauge nodes and bifundamental hypermultiplets as the quiver for the $+_{N,M}$ theory in (\ref{eq:D5NS5-quiver}).
However, the flavor hypermultiplets at the boundary nodes of the respective quivers are different, $N$ fundamental hypermultiplets for the $+_{N,M}$ theory compared to one antisymmetric hypermultiplet at each end for the $\ZZ_{2k}$ orbifold of the $USp(N)$ theory, and the length $k$ of the quiver in \cite{Jafferis:2012iv} is order one.
For the quivers in \cite{Jafferis:2012iv}, saddle points were found with equal eigenvalue distributions for all gauge nodes. 
For the $+_{N,M}$ theories, on the other hand, the saddle point configuration (\ref{eq:rho-sol-plus}) depends non-trivially on the gauge node label $z$.
For the free energy this leads to different scalings: 
$N^2M^2$ for the $+_{N,M}$ theory compared to $N^{5/2}k^{3/2}$ for the orbifolds of the $USp(N)$ theory.

\subsection{\texorpdfstring{$T_N$}{T[N]} theory}\label{sec:TN}

The (unconstrained) $T_N$ theory is defined by a junction of $N$ D5, $N$ NS5 and $N$ $(1,1)$ 5-branes \cite{Benini:2009gi}, as shown in fig.~\ref{fig:TN}.
It is the strongly-coupled UV fixed point of the linear quiver gauge theory \cite{Bergman:2014kza,Hayashi:2014hfa}
\begin{align}
\label{TNquiver}
[2]-(2)-(3)-\ldots-(N-2)-(N-1)-[N] ~,
\end{align}
with all Chern-Simons levels zero.
The S-dual deformation leads to the same quiver.
For $N=3$ this is the rank-1 $E_6$ theory.
The $S^5$ free energy at large $N$ has been obtained numerically from localization in \cite{Fluder:2018chf}, and matched to an analytic supergravity prediction from \cite{Gutperle:2017tjo}. For the large-$N$ limit of the matrix model, one can strictly speaking not expect $\hat\rho_t$ to be a smooth distribution for small $t$ where $SU(t+1)$ has small rank. 
But one can nevertheless use it as an approximation, which will lead to consistent results.
In (\ref{eq:cF-gen-3}) the $T_N$ quiver corresponds to $L=N-2$ and $N_t=t+1$, and the only non-vanishing $k_t$ are $k_1=2$ and $k_{N-2}=N$. In the continuous version (\ref{eq:cF-gen-6}),
\begin{align}\label{eq:TN-data}
N(z)&=Nz~, &k(z)&=\frac{2}{L}\delta(z)+\frac{N}{L}\delta(z-1)~,
&
c(z)&=0~.
\end{align}
This theory has $N_f=2N$ at all interior gauge nodes, the only exception is the boundary node at the left end of the quiver tail, where $N(z)$ vanishes. Thus, the Lagrange multiplier functions $\mu(z)$ and $\tau(z)$ in (\ref{eq:cF-gen-6}) can again be set to zero.
The expression for $\cF$ in eq.~(\ref{eq:cF-gen-6}) becomes
\begin{align}\label{eqn:cFTN-6}
\cF_{T_N}  \ = \  
&
N^2\int dz\,dx\,dy\,\cL
+N^5\int dx\, \hat\rho(1,x)\left[F_H(x)-\frac{1}{2}\int dy\,\hat\rho(z,y)F_H(x-y)\right]\,.
\end{align}
The two flavors at the $SU(2)$ gauge node only produce subleading contributions; they drop out in the large $N$ limit due to $N(0)=0$.
The saddle point eigenvalue distribution is given by (\ref{eq:varrho-s}) with the quiver data (\ref{eq:TN-data}), 
\begin{align}\label{eq:TN-saddle}
 \hat\rho_s(z,x)&=\frac{\sin (\pi  z)}{z}\frac{1}{\cosh \left(2\pi x\right)+\cos (\pi  z)}~.
\end{align}
The free energy is obtained from (\ref{eq:free-energy-gen}) with (\ref{eq:cF-gen-Nf2NC}) and (\ref{eq:TN-data}), which yields
\begin{align}\label{eqn:cFTN-eval-4}
F_{\vec{\omega}}^{T_N} & = 
-\frac{1}{8\pi^2}\frac{\omega_{\rm tot}^3}{\omega_1\omega_2\omega_3}\zeta(3)N^4~.
\end{align}
For the round $S^5$ with $\vec{\omega}=(1,1,1)$ this provides an analytic result matching the field theory numerics of \cite{Fluder:2018chf} and the supergravity computations of \cite{Gutperle:2017tjo}. 
The two flavors at the left end of the quiver (\ref{TNquiver}) did not explicitly play a role in the derivation, but regularity of the eigenvalue distribution at the quiver tail did.

\subsection{\texorpdfstring{$Y_N$}{Y[N]} theory}
The $Y_N$ theories were defined in \cite{Bergman:2018hin} on junctions of $N$ $(1,1)$ 5-branes, $N$ $(-1,1)$ 5-branes and $2N$ NS5-branes, as shown in fig.~\ref{fig:YN}. The theory admits two quiver deformations that were discussed in \cite{Bergman:2018hin}, and we compute the free energy from both of them.

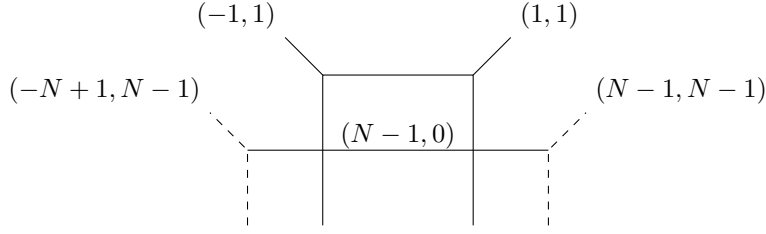
\begin{figure}
\begin{tikzpicture}
 \draw (-2,0) -- (2,0);

 \draw (-1,-1) -- (-1,1) -- (1,1) -- (1,-1);
 \draw (1,1) -- +(0.5,0.5) node [anchor=south west] {\small $(1,1)$};
 \draw (-1,1) -- +(-0.5,0.5) node [anchor=south east] {\small $(-1,1)$};
 \node at (0,0.2) {\small $(N-1,0)$};
 
 \draw[dashed] (-2,-1) -- (-2,0) -- (-2.5,0.5) node [anchor=south east] {\small $(-N+1,N-1)$};
 \draw[dashed] (2,-1) -- (2,0) -- (2.5,0.5) node [anchor=south west] {\small $(N-1,N-1)$};
\end{tikzpicture}
\caption{
$Y_N$ junction with the central node of the quiver deformation (\ref{eq:YN-quiver-2}) partly resolved. 
The solid lines show the subweb correspobding to the central node; it can be obtained from a $+_{N,2}$ web by integrating out two flavors. The quiver tails correspond to the dashed lines.
\label{fig:YN-central}}
\end{figure}

The quiver gauge theory obtained directly from the $Y$-shaped 5-brane junction reads
\begin{align}\label{eq:YN-quiver-2}
[2]-(2)-(3)-\cdots-(N-1)-(N)_{\pm 1}-(N-1)-\cdots-(3)-(2)-[2] \,. 
\end{align}
Along the two quiver tails, $N_f=2N$ for each node, and the Chern-Simons levels are zero.
At the central node $N_f=2(N-1)$, so the quantization condition $c_{\rm cl}+\frac{1}{2}N_f\in\ZZ$ \cite{Intriligator:1997pq} requires an integer Chern-Simons level.
The brane web realization of the central node is shown in fig.~\ref{fig:YN-central}.
It can be obtained from a $+_{2,N}$ web, which has Chern-Simons level zero, by integrating out two flavors. 
The Chern-Simons level at the central node therefore is $\pm 1$.
The quiver is also related to the quiver for the $T_{2K,K,2}$ theory in (\ref{eq:TNKj-quiver-2}) below by replacing the two flavors at the central node by the Chern-Simons term.

The data describing the quiver (\ref{eq:YN-quiver-2}) in (\ref{eq:quiver}) is $L=2N-3$, $N_t=t+1$ for $1\leq t \leq N-1$ and $N_t=2N-t-1$ for $N\leq t\leq L$. The non-vanishing $k_t$ are $k_1=k_L=2$.
The only non-vanishing Chern-Simons level is $c_{N-1}=\pm 1$.
The continuous version is
\begin{align}
 N(z)&=N\begin{cases} 
            2z & z\leq \frac{1}{2}\\
            2-2z &z\geq \frac{1}{2}
        \end{cases}
~,
&
k(z)&=\frac{2}{L}\delta(z)+\frac{2}{L}\delta(z-1)~,
\nonumber\\
&&c(z)&=\frac{c_{N-1}}{L}\delta(z-\tfrac{1}{2})~.
\end{align}
Since $\mu(z)$ and $\tau(z)$ vanish along parts of the quiver where $N_f=2N$, they take the form
\begin{align}
 \mu(z)&=\frac{\mu_0}{L}\delta\big(z-\tfrac{1}{2}\big)~, & \tau(z)&=\frac{\tau_0}{L}\delta\big(z-\tfrac{1}{2}\big)~.
\end{align}
In the expression for $\cF$ in (\ref{eq:cF-gen-6}) the flavors at the boundary nodes drop out, and the boundary terms vanish, such that
\begin{align}\label{eq:cF-YN}
 \cF_{Y_N} \ = \ & L^2\int_0^1 dz \int dx\,dy\, \cL
 +\frac{\pi}{3}c_{N-1}L^3N\int dx\,\hat\rho(\tfrac{1}{2},x) x^3
 \nonumber\\&
 +L^3N\left[\mu_0\int dx\,x\hat\rho(\tfrac{1}{2},x)+\tau_0\left(\int dx\,\hat\rho(\tfrac{1}{2},x)-1\right)\right]
 ~.
\end{align}

\subsubsection{Saddle point}

The quiver (\ref{eq:YN-quiver-2}) is symmetric under $z\rightarrow 1-z$, as reflected in $N(z)=N(1-z)$, and the same is expected for the saddle point configuration.
One therefore has to find a non-negative harmonic function $\varrho(z,x)$ on the strip $(z,x)\in [0,\tfrac{1}{2}]\times\RR$, with,
since $N(z)$ vanishes at $z=0$,
\begin{align}\label{eq:YN-bc-1}
 \varrho(0,x)&=0~.
\end{align}
The boundary condition at $z=\tfrac{1}{2}$ follows from the junction condition in (\ref{eq:junction-gen}), which with the symmetry under $z\rightarrow 1-z$ becomes
\begin{align}\label{eq:YN-bc}
 -2\int dy\,\partial_z\varrho(z,y)\big\vert_{z=z_0-\epsilon}F_H(x-y)
 +\frac{\pi}{3}Lc_{N-1} x^3+L\mu_0+L\tau_0 x&=0~.
\end{align}
Since $2L|c_{N-1}|$ equals the discontinuity in $\partial_z N(z)$ at $z=\tfrac{1}{2}$, 
following (\ref{eq:junction-constr}) the support at $z=\frac{1}{2}$ has to be restricted to 
$x<x_0$ for $c_{N-1}=+1$ and to $x>x_0$ for $c_{N-1}=-1$.
Consequently, (\ref{eq:YN-bc}) has to hold for $c_{N-1}(x-x_0)<0$. 
Acting on (\ref{eq:YN-bc}) with $(\partial_x)^4$ shows that $\partial_z\varrho(z,x)\vert_{z=\frac{1}{2}}=0$ for $c_{N-1}(x-x_0)<0$.
Thus, the boundary conditions at $z=\frac{1}{2}$ are
\begin{align}
 \varrho\big(\tfrac{1}{2},x\big)&=0 \qquad \text{for $c_{N-1}(x-x_0)>0$}~, 
 \nonumber \\
 \partial_z\varrho\big(z,x\big)\vert_{z=\tfrac{1}{2}}&=0 \qquad \text{for $c_{N-1}(x-x_0)<0$}~.
 \label{eq:YN-bc-2}
\end{align}

To construct $\varrho$, the half of the strip with $z\in[0,\tfrac{1}{2}]$ is mapped to the upper half plane with complex coordinate $u$ via
\begin{align}\label{eq:YN-u}
 u&=e^{-4\pi c_{N-1}(x-x_0)+2\pi i z}~.
\end{align}
The range $c_{N-1}(x-x_0)< 0$ at $z=\frac{1}{2}$ is mapped to $(-\infty,-1)$ on the real line,
while $c_{N-1}(x-x_0)> 0$ at $z=\frac{1}{2}$ is mapped to $(-1,0)$ on the real line. 
The boundary at $z=0$ is mapped to the positive real line.
The boundary conditions (\ref{eq:YN-bc-1}), (\ref{eq:YN-bc-2}) thus require $\varrho(u)$ to satisfy 
Neumann boundary condition on $(-\infty,-1)$ and vanishing Dirichlet on $(-1,\infty)$.
Moreover, $\varrho$ should have at most an integrable divergence at $u=-1$ such that the eigenvalue distribution can have finite norm,
and it should vanish as $u\rightarrow\infty$. 
The entire strip with $z\in [0,1]$ maps to the entire complex plane, with the same conditions imposed on the real line.
Since $c_{N-1}(x-x_0)< 0$ at $z=\tfrac{1}{2}$ corresponds to $(-\infty,-1)\subset\RR$, $\varrho$ should be smooth across that part of the real line, but may have a branch cut from $-1$ along the positive real axis.
Such a function is readily found as 
\begin{align}
 \varrho(u)&=\frac{2N}{\sqrt{-u-1}}+\rm{c.c.}
\end{align}
with the branch cut of the square root $\sqrt{\cdot}$ along the negative real axis, such that the branch cut of $\varrho$ extends from $u=-1$ along the positive real axis.
The overall real coefficient has been fixed by demanding proper normalization of the eigenvalue distributions.
The remaining parameter $x_0$ implicit in the definition of $u$ in (\ref{eq:YN-u})  is determined by the $SU(N)$ constraint, which yields
\begin{align}
 x_0&=\frac{c_{N-1}}{2\pi}\ln 2~.
\end{align}
The final result for the saddle point configuration for $z\in[0,1]$ therefore is
\begin{align}\label{eq:varrho-YN}
 \hat\rho_s(z,x)&=\frac{2N}{N(z)\sqrt{-1-4e^{-4\pi c_{N-1}x+2\pi iz}}}+\rm{c.c.}
\end{align}

It remains to verify that the condition (\ref{eq:YN-bc}) is satisfied, and determine $\mu_0$ and $\tau_0$ which will be needed for computing the free energy.
The first term in (\ref{eq:YN-bc}),
\begin{align}\label{eq:YN-bc-2-a}
 T(x)&\equiv\int dy\, |x-y|^3\partial_z\varrho_s(z,y)\Big\vert_{z=\tfrac{1}{2}-\epsilon}~,
\end{align}
is a polynomial of degree $4$ in $x$: Acting with $(\partial_x)^n$ with $n\geq 5$ leaves derivatives of $\delta$-functions in the integral. Using integration by parts and the fall-off behavior of $\varrho$ shows that these derivatives vanish.
Thus,
\begin{align}
 T(x)&=\sum_{n=0}^4 a_n x^n~,
 &
 a_n&=\frac{1}{n!}(\partial_x)^nT(x)\Big\vert_{x=0}~.
\end{align}
The coefficients can be evaluated using the support properties of $\partial_z\varrho$ at $z=\frac{1}{2}$.
This shows that (\ref{eq:YN-bc}) is indeed satisfied with
\begin{align}\label{eq:YN-mu0-tau0}
 \mu_0&=\frac{\zeta(3)}{16\pi^2}~,
 &
 \tau_0&=-\frac{\pi}{48}c_{N-1}~.
\end{align}

\subsubsection{Free energy}
To derive the free energy, the expression for $\cF_{Y_N}$ in (\ref{eq:cF-YN}) is evaluated on the saddle point configuration (\ref{eq:varrho-YN}). The Lagrange multiplier terms do not contribute since they multiply the constraints, leaving
\begin{align}\label{eq:cF-YN-eval}
 \cF_{Y_N}\big\vert_{\hat\rho=\hat\rho_s} \ = \ & L^2\int_0^{1} dz \int dx\,dy\, \cL\big\vert_{\hat\rho=\hat\rho_s}
 +\frac{\pi}{3}c_{N-1}L^3N\int dx\,\hat\rho_s(\tfrac{1}{2},x) x^3
 ~.
\end{align}
Using integration by parts and that  $\varrho_s$ is harmonic, the first term reduces to boundary terms. 
Using also the symmetry of $\varrho_s$ under $z\rightarrow 1-z$, we find
\begin{align}
 \cF_{Y_N}\big\vert_{\hat\rho=\hat\rho_s} &=
 L^2\int dx\,\varrho_s(\tfrac{1}{2},x)\left[\frac{\pi}{3}c_{N-1}Lx^3 - \int dy\,\partial_z\varrho_s(z,y)\big\vert_{z=\frac{1}{2}-\epsilon}F_H(x-y)\right]
 ~. 
\end{align}
With the junction condition (\ref{eq:YN-bc}) this becomes
\begin{align}
 \cF_{Y_N}\big\vert_{\hat\rho=\hat\rho_s} &=
 L^3\int dx\,\varrho_s(\tfrac{1}{2},x)\left[\frac{\pi}{6}c_{N-1}x^3 - \frac{\mu_0}{2}-\frac{\tau_0}{2}x\right]
 ~. 
\end{align}
The term proportional to $\tau_0$ vanishes by virtue of the $SU(N)$ constraint, 
the one with $\mu_0$ can be evaluated using the normalization of $\varrho_s$.
With (\ref{eq:YN-mu0-tau0}) and (\ref{eq:free-energy-gen}) one finds
\begin{align}\label{eq:FS5-YN}
 F_{\vec{\omega}}^{Y_N}&=-\frac{1}{2\pi^2}\frac{\omega_{\rm tot}^3}{\omega_1\omega_2\omega_3}\zeta(3)N^4~.
\end{align}
This free energy is related to that of the $T_N$ theory by a factor $4$. This relation becomes more transparent in the S-dual quiver deformation, which will be discussed next.

\subsubsection{S-dual quiver}
The quiver deformation arising after performing an S-duality on the brane web is given by
\begin{align}
(2)-(4)-(6)-\ldots-(2N-2)-[2N] ~,
\end{align}
with all Chern-Simons levels zero and $N_f=2N$ at all nodes.
For $N=2$ this is the rank-1 $E_5$ theory.
In the matrix model (\ref{eq:cF-gen-3}) this quiver corresponds to $L=N-1$, $N_t=2t$ and $k_{N-1}=2N$.
The continuous version (\ref{eq:cF-gen-6}) is specified by
\begin{align}
 N(z)&=2N z~, & k(z)&=\frac{2N}{L}\delta(z-1)~, & c(z)&=0~.
\end{align}
Since $N_f=2N$ at all nodes, $\mu(z)=\tau(z)=0$.
Consequently, $\cF$ in (\ref{eq:cF-gen-6}) becomes
\begin{align}
 \cF_{Y_N}&=N^2\int_0^1 dz\int dx\,dy\,\cL +2N^5\int dx\,\hat\rho(1,x)\left[2F_H(x)- \int dy\,\hat\rho(1,y)F_H(x-y)\right]~.
\end{align}
Up to an overall factor of four this is equivalent to $\cF_{T_N}$ for the $T_N$ theory in (\ref{eqn:cFTN-6}), 
where the $[2]$ fundamentals in the quiver (\ref{TNquiver}) only produce subleading corrections. 
The saddle point conditions are insensitive to this overall factor.
Consequently, the free energy for the $Y_N$ theory is related to that of the $T_N$ theory in (\ref{eqn:cFTN-eval-4}) by a factor $4$,
$F_{\vec{\omega}}^{Y_N}=4F_{\vec{\omega}}^{T_N}$, leading to (\ref{eq:FS5-YN}).
From the supergravity perspective this relation between the $T_N$ and $Y_N$ theories at large $N$ follows from the discussion of combined $SL(2,\RR)$ transformations and overall rescaling of the 5-brane charges in sec.~4 of \cite{Bergman:2018hin}.

\subsection{\texorpdfstring{$\pslash_N$}{pslash} theory}

The $\pslash_N$ theory was defined in \cite{Bergman:2018hin} on a sextic intersection of $N$ D5-branes, $N$ NS5-branes and $N$ $(1,1)$ 5-branes, as shown in fig.~\ref{fig:pslash}. 
It describes the strongly-coupled UV fixed point of the quiver gauge theory
\begin{align}\label{eq:pslash-quiver}
 [N]-(N+1)-\ldots - (2N-1)-(2N)-(2N-1)-\ldots -(N+1)-[N]~,
\end{align}
with all Chern-Simons levels zero (the subweb corresponding to the central node is symmetric under rotation by $\pi$, corresponding to charge conjugation). 
For $N=1$ this is the rank-1 $E_3$ theory.
The data characterizing this theory in (\ref{eq:cF-gen-3}) is $L=2N-1$ and $N_t=N+t$ for $t\leq N$ while $N_t=3N-t$ for $t\geq N$.
The non-zero $k_t$ are $k_1=k_{2N-1}=N$. 
The continuous version in (\ref{eq:cF-gen-6}) is defined by
\begin{align}
 N(z)&=N\begin{cases} 1+2z~, & z\leq \tfrac{1}{2}\\ 3-2z~,& z\geq \frac{1}{2}\end{cases}~,
 &
 k(z)&=\frac{N}{L}\left(\delta(z)+\delta(z-1)\right)~,
 &
 c(z)&=0~.
\end{align}
Since the Lagrange multipliers vanish along parts of the quiver where $N_f=2N$, their form is
\begin{align}
 \mu(z)&=\frac{\mu_0}{L}\delta\big(z-\tfrac{1}{2}\big)~, & \tau(z)&=\frac{\tau_0}{L}\delta\big(z-\tfrac{1}{2}\big)~.
\end{align}
With this data $\cF$ in eq.~(\ref{eq:cF-gen-6}) for the $\pslash_{N}$ theory becomes
\begin{align}\label{eq:cF-pslash}
 \cF_{\pslash_N} \ = \ &
 4N^2\!\int_0^1 \!dz\! \int \!dx\,dy\, \cL
+4N^5\!\!\!\sum_{z\in\lbrace 0,1\rbrace}\int dx\, \hat\rho(z,x)\left[F_H(x)-\frac{1}{2}\int \!dy\,\hat\rho(z,y)F_H(x-y)\right]
\nonumber\\&
+2L^3N\left[\mu_0\int dx\,x\hat\rho(\tfrac{1}{2},x)+\tau_0\left(\int dx\,\hat\rho(\tfrac{1}{2},x)-1\right)\right].
\end{align}

The quiver is symmetric under $z\rightarrow 1-z$ and the saddle point eigenvalue distributions are expected to be symmetric as well.
We therefore have to construct a harmonic function on $(z,x)\in [0,\tfrac{1}{2}]\times\RR$. 
The boundary condition at $z=0$ reads
\begin{align}\label{eq:pslash-bc-1}
 \varrho(0,x)&=N\delta(x)~.
\end{align}
The boundary condition at $z=\tfrac{1}{2}$ follows from the junction condition (\ref{eq:junction-gen}). 
With the symmetry under $z\rightarrow 1-z$ it becomes
\begin{align}\label{eq:pslash-bc-2}
  2\int dy\,\partial_z\varrho(z,y)\big\vert_{z=\frac{1}{2}-\epsilon}F_H(x-y)
  &=L(\mu_0+\tau_0 x)~.
\end{align}
Since the Chern-Simons level is zero, the support of the eigenvalue distribution at $z=\frac{1}{2}$ has to be bounded from below and from above.
Since, with no Chern-Simons terms, the problem is symmetric under $x\rightarrow -x$,
the boundary conditions at $z=\frac{1}{2}$ are
\begin{align}
 \varrho\big(\tfrac{1}{2},x\big)&=0\quad \text{for $|x|>x_0$}~,
 \nonumber\\
 \partial_z\varrho(z,x)\big\vert_{z=\frac{1}{2}}&=0\quad \text{for $|x|<x_0$}~.
\end{align}

\subsubsection{Saddle point}

To construct $\varrho$, the problem is mapped to the upper half plane with a following $SL(2,\RR)$ transformation,
\begin{align}
 u&=e^{4\pi x+2\pi i z}~,
 &
 v&=\frac{u e^{4 \pi x_0}+1}{u+e^{4 \pi x_0}}~.
\end{align}
In the $v$ coordinate we need a non-negative function satisfying Neumann boundary conditions for 
$v$ in $(-\infty,0)\subset\RR$ and Dirichlet boundary conditions for $v\in\RR^+$, 
\begin{align}
\varrho(v)\big\vert_{v\in\RR^+}=N\delta(v,1)~. 
\end{align}
With this condition the eigenvalue distributions vanish for $x\rightarrow \pm \infty$, as required for normalizability.
The additional requirements are the following: There should be at most integrable divergences at $v=0$ and $v=\infty$, for normalizable eigenvalue distributions. Aside from the $\delta$-function pole at $v=1$ these should be the only divergences.
The entire strip is mapped to the entire complex plane, and $\varrho$ should be smooth across the negative real axis.
Moreover, the eigenvalue distributions should be symmetric under $x\rightarrow -x$, i.e.\ $v\rightarrow 1/\bar v$.

The function $\varrho$ is constructed in two steps. 
A function satisfying the specified boundary conditions, symmetry under $v\rightarrow 1/\bar v$ and the remaining requirements is given by
\begin{align}
 \varrho^{}_1&=\frac{a\sqrt{-v}}{1-v}+\mathrm{c.c.}
\end{align}
with the branch cut of the square root along the negative real axis, such that $\varrho^{}_1$ has the branch cut along the positive real axis.
We may add an arbitrary function satisfying vanishing Dirichlet boundary conditions on all of $\RR^+$ and all other requirements.
Such a function is given by
\begin{align}
 \varrho^{}_0&=\frac{b(1-v)}{\sqrt{-v}}+\mathrm{c.c.}
\end{align}
It has square root divergences at the origin and at the point at infinity, and the relative coefficients of the terms in the numerator are fixed by the requirement for invariance under $v\rightarrow 1/\bar v$.

The $SU(N)$ constraint is satisfied automatically due to the symmetry under $x\rightarrow -x$,
and the parameters $a$ and $b$ are fixed by the normalization conditions,
\begin{align}
 a&=2N\tanh(2\pi x_0)~,
 &
 b&=\frac{1}{2} N\coth (\pi x_0) \sech(2 \pi x_0)~.
\end{align}
Finally, $x_0$ is determined from the junction condition (\ref{eq:pslash-bc-2}). 
The left hand side is a polynomial of degree $4$ in $x$, by the same argument as for the $Y_N$ theory,
\begin{align}\label{eq:pslash-bc-2-a}
 T(x)&\equiv\int dy\, |x-y|^3\partial_z\varrho_s(z,y)\Big\vert_{z=\tfrac{1}{2}-\epsilon}
 =\sum_{n=0}^4 a_n x^n~,
 &
 a_n&=\frac{1}{n!}(\partial_x)^nT(x)\Big\vert_{x=0}~.
\end{align}
The quartic term vanishes due to the Neumann boundary condition on $(-x_0,x_0)$ at $z=\frac{1}{2}$.
The linear and cubic terms vanish by symmetry of $\varrho_s$ under $x\rightarrow -x$.
The condition that the quadratic term be zero leads to
\begin{align}\label{eq:pslash-x0}
 \cosh(2\pi x_0)&=2~.
\end{align}
The resulting saddle point configuration for $z\in(0,1)$ is given by
\begin{align}\label{eq:pslash-saddle}
 \hat\rho_s&=
 \frac{N}{N(z)}\frac{1-2 \csch^2(2\pi x+i\pi z)}{\sqrt{3}+2 \coth(2\pi x+i\pi z)}\sqrt{\frac{\sqrt{3} \tanh(2\pi x+i\pi z)+2}{\sqrt{3} \tanh(2\pi x+i\pi z)-2}}
 +\mathrm{c.c.}
\end{align}
The junction condition (\ref{eq:pslash-bc-2}) is satisfied with 
\begin{align}\label{eq:pslash-mu0}
 \mu_0&=\frac{7\zeta(3)N}{12\pi^2L}~,
 &
 \tau_0=0~.
\end{align}

\subsubsection{Free energy}
The free energy is obtained by evaluating $\cF$ in (\ref{eq:cF-pslash}) on the saddle point configuration (\ref{eq:pslash-saddle}).
With the boundary conditions at $z=0$ and $z=1$, the local saddle point equation and the symmetry under $z\rightarrow 1-z$ this leads to
\begin{align}\label{eq:cF-pslash-rhos}
 \cF_{\pslash_N}\big\vert_{\varrho=\varrho_s}
 &=-4N^2\int dx\, dy\,\left[\varrho_s(z,x)\partial_z\varrho_s(z,y)\right]_{z=0}^{z=\frac{1}{2}-\epsilon}F_H(x-y)~.
\end{align}
Using the boundary condition at $z=0$ this further evaluates to
\begin{align}
 \cF_{\pslash_N}\big\vert_{\varrho=\varrho_s}
 &=
 4N^2\!\int \!dx\,\Big[N\partial_z\varrho_s(z,x)\big\vert_{z=0}F_H(x)
 -\varrho_s(\tfrac{1}{2},x)\int dy\,\partial_z\varrho_s(z,y)\big\vert_{z=\frac{1}{2}-\epsilon}F_H(x-y)\Big].
\end{align}
The remaining integral in the second term can be evaluated using (\ref{eq:pslash-bc-2}) with (\ref{eq:pslash-mu0})
and the normalization of $\varrho_s$, which yields
\begin{align}
 \cF_{\pslash_N}\big\vert_{\varrho=\varrho_s}
 &=
 4N^3\left[\int dx\,\partial_z\varrho_s(z,x)\big\vert_{z=0}F_H(x) - L\mu_0\right]
 =
 -\frac{7}{2\pi^2}\zeta(3)N^4~.
\end{align}
The resulting free energy is
\begin{align}\label{eq:F-pslash}
 F_{\pslash_N}& =-\frac{7}{2 \pi ^2} \frac{\omega_{\rm tot}^3}{\omega_1\omega_2\omega_3}\zeta (3) N^4 ~.
\end{align}
For the round sphere this agrees with a supergravity computation of the same quantity along the lines of \cite{Gutperle:2017tjo}.
As in the previous examples the result involves an overall $\zeta(3)$ and has a simple dependence on the parameters characterizing the field theory.
Examples where the free energy has more complicated dependence on the parameters are discussed in the following sections.

\begin{figure}
 \subfigure[][]{\label{fig:T2K-K-2}
 \begin{tikzpicture}[scale=0.95]
    \foreach \i in {-3/2,-1/2,1/2,3/2}{
      \draw (-0.4,0.127*\i) -- (-1.8,0.127*\i);
      \draw (0.127*\i,-0.4) -- (0.127*\i,-1.8) [fill=black] circle (1pt);
      \draw (0.28+0.09*\i,0.28-0.09*\i) -- (1.27+0.09*\i,1.27-0.09*\i) [fill=black] circle (1pt) ;
    }
    \draw[fill=black] (-1.8,0.127) ellipse (1pt and 2.8pt);
    \draw[fill=black] (-1.8,-0.127) ellipse (1pt and 2.8pt);
    \draw[fill=gray] (0,0) circle (0.38);
    
    \node [anchor=east] at (-1.9,0) {\footnotesize $[K,K]$};
    \node [anchor=south west] at (1.3,1.3) {\footnotesize $2K$};
    \node [anchor=north] at (0,-1.9) {\footnotesize $2K$};
\end{tikzpicture}
}\hskip 0mm
 \subfigure[][]{\label{fig:TNKj}
 \begin{tikzpicture}[scale=0.95]
    \foreach \i in {-2,-1,0,1,2}{
      \draw (-0.4,0.127*\i) -- (-1.8,0.127*\i);
      \draw (0.127*\i,-0.4) -- (0.127*\i,-1.8) [fill=black] circle (1pt);
      \draw (0.28+0.09*\i,0.28-0.09*\i) -- (1.27+0.09*\i,1.27-0.09*\i) [fill=black] circle (1pt) ;
    }
    \draw (-1.8,0.127*2) [fill=black] circle (1pt);
    \draw (-1.8,0.127*1) [fill=black] circle (1pt);
    \draw[fill=gray] (0,0) circle (0.38);
    \draw[fill=black] (-1.8,-0.127*1) ellipse (1pt and 4.4pt);

    \node [anchor=east] at (-1.9,0) {\footnotesize $[K^j,N-jK]$};
    \node [anchor=south west] at (1.3,1.3) {\footnotesize $N$};
    \node [anchor=north] at (0,-1.9) {\footnotesize $N$};
\end{tikzpicture}
}\hskip 8mm
\subfigure[][]{\label{fig:plus-NMk}
  \begin{tikzpicture}[scale=0.75]
  \draw[fill=gray] (0,0) circle (0.65);
  \foreach \i in {-5,-3,-1,1,3,5}{
   \draw (0.8,0-0.08*\i) -- +(1.4,0) [fill];
   \draw (-0.8,0-0.08*\i) -- +(-1.4,0) [fill] circle (1.3pt);
  }
  \draw[fill] (0.8+1.4,0.08*3) ellipse (1.3pt and 5.5pt);
  \draw[fill] (0.8+1.4,-0.08*3) ellipse (1.3pt and 5.5pt);

  \foreach \i in {-6,-3,0,3,6}{
   \draw (0.06*\i,0.8) -- +(0,1.4) [fill] circle (1.3pt);
   \draw (0.06*\i,-0.8) -- +(0,-1.4) [fill] circle (1.3pt);
  }
  \node at (0.8+1.75,0) {\scriptsize $j$};
  \node at (-0.8-1.75,0) {\scriptsize $N$};
  \node at (0,0.8+1.75) {\scriptsize $M$};
  \node at (0,-0.8-1.75) {\scriptsize $M$};
  \end{tikzpicture}
}
\caption{Constrained 5-brane junctions with multiple 5-branes ending on the same 7-branes.
From left to right for the $T_{2K,K,2}$, $T_{N,K,j}$, and $+_{N,M,j}$ theories.\label{fig:constrained}}
\end{figure}
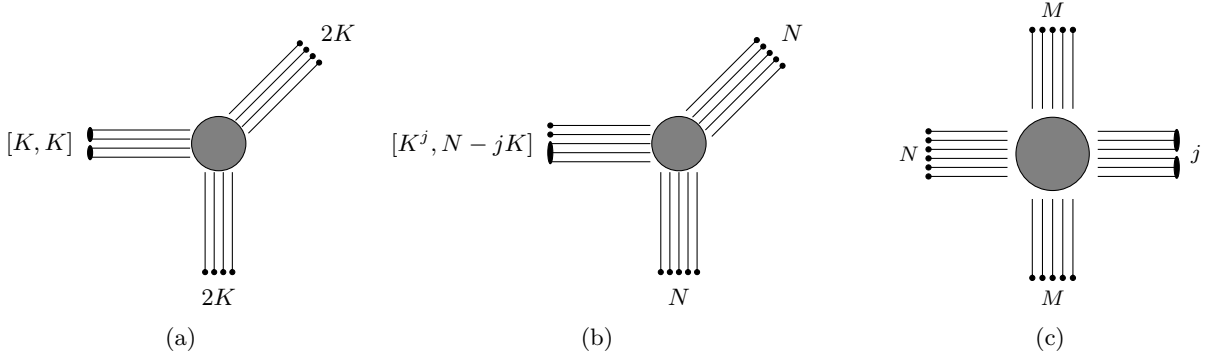

\subsection{\texorpdfstring{$T_{2K,K,2}$}{T[K,K/2,2]} theory}\label{sec:T2K-K-2}

The $T_{2K,K,2}$ theories, as defined in \cite{Chaney:2018gjc}, are realized in Type IIB string theory by junctions involving the same 5-branes as the unconstrained $T_N$ theories of sec.~\ref{sec:TN}, but with the D5 branes partitioned into two groups of $K$ D5 branes, with each group ending on one D7-brane, fig.~\ref{fig:T2K-K-2}. 
These theories are obtained from the $T_N$ theories with $N=2K$ by renormalization group flows.
The supergravity duals and aspects of the field theories were discussed in \cite{Chaney:2018gjc}.
They may also be defined as UV fixed points of the linear quiver gauge theories
\begin{align}\label{eq:TNKj-quiver-2}
 [2]-(2)-(3)-\ldots - (K-1)- (&K) - (K-1) -\ldots -(3)-(2)-[2] ~,\nonumber\\
 &\,|\, \\
 &\,\!\![2]\nonumber
\end{align}
with all Chern-Simons levels zero.
For $K=2$ this is the rank-1 $E_7$ theory.
The $S^5$ free energy was obtained numerically from localization and in supergravity in \cite{Fluder:2019szh}, and matched to very good accuracy between the two descriptions.

In (\ref{eq:cF-gen-3}), the $T_{2K,K,2}$ quiver corresponds to $L=2K-3$ with $N_t=t+1$ for $t\leq K-1$ and $N_t=2K-t-1$ for $t\geq K$.
The non-vanishing $k_t$ are $k_1=k_{K-1}=k_{2K-2}=2$. The continuous version in (\ref{eq:cF-gen-6}) is defined by
\begin{align}\label{eq:T-2KK2-data}
 N(z)&=\begin{cases}
         2Kz \hskip 17mm z<\frac{1}{2}
         \\
         2K(1-z) \qquad z>\frac{1}{2}
       \end{cases},
 &
 k(z)&=\frac{2}{L}\left(\delta(z)+\delta\big(z-\tfrac{1}{2}\big)+\delta(z-1)\right)~,
\end{align}
with $c(z)=0$.
Since $N_f=2N$ at all interior nodes, $\mu(z)=\tau(z)=0$.
In (\ref{eq:cF-gen-6}) the flavors at the $SU(2)$ nodes produce subleading contributions only and drop out.
Thus,
\begin{align}
 \label{eq:F-Tj2-1}
 \cF_{T_{2K,K,2}} \ = \ & 
 4K^2\int_0^1 dz\int dx\,dy\,\cL  
 + 16K^4\int dx\,  \hat\rho\Big(\frac{1}{2},x\Big)F_H(x)\, .
\end{align}
The quiver is symmetric under reflection across the central node, and the saddle point eigenvalue distribution is expected to be symmetric as well.
Since $N_f=2N$ at all interior nodes, the discussion of sec.~\ref{sec:Nf2NC-sol} applies, and the saddle point is given by (\ref{eq:varrho-s}) with the data in (\ref{eq:T-2KK2-data}),
\begin{align}\label{eq:T-2K-K-2-saddle}
\hat\rho_s(z,x)&=\frac{2K}{\pi N(z)}
\ln\left[\frac{\cosh\left(2\pi x\right)+\sin\left(\pi z\right)}{\cosh\left(2\pi x\right)-\sin\left(\pi z\right)}\right]~.
\end{align}
The free energy is obtained via (\ref{eq:free-energy-gen}) with (\ref{eq:cF-gen-Nf2NC}) and (\ref{eq:T-2KK2-data}), which yields
\begin{align}\label{eq:Tj2-result}
 F_{\vec{\omega}}^{T_{2K,K,2}}  &=
 -\frac{31}{4\pi^4}\frac{\omega_{\rm tot}^3}{\omega_1\omega_2\omega_3}\zeta(5) K^4~.
\end{align}
For the round sphere with $\vec{\omega}=(1,1,1)$, this is the analytic representation of the numerical field theory and supergravity results of \cite{Fluder:2019szh}.

From the supergravity perspective the appearance of $\zeta(5)$ instead of $\zeta(3)$  may be understood as follows.
The general $AdS_6$ solutions providing the holographic duals for 5-brane junctions are defined by a pair of locally holomorphic functions $\cA_\pm$.
These functions have meromorphic differentials for solutions corresponding to unconstrained junctions \cite{DHoker:2017mds},
while the differentials involve logarithms for solutions corresponding to constrained junctions \cite{DHoker:2017zwj}.
In the convention of \cite{Goncharov:2010jf}, $\cA_\pm$ as functions of the poles have transcendentality degree one for solutions without 7-branes and two for solutions with 7-branes.
The free energies are computed from certain integrals of the functions $\cA_\pm$, which are thus of higher transcendentality degree for solutions corresponding to constrained junctions. 
From the field theory perspective the appearance of $\zeta(5)$ is an effect of the charges due to fundamental flavors at internal nodes, as seen explicitly from (\ref{eq:cF-gen-Nf2NC}).

\subsection{\texorpdfstring{$T_{N,K,j}$}{T[N,K,j]} theory}

The 5-brane realization of the $T_{N,K,j}$ theories is obtained from the one for the $T_N$ theories by taking the $N$ D5-branes and separating out $j$ groups of $K$ D5-branes that each end on a single D7 brane. This leaves $N-jK$ unconstrained D5-branes, as shown in fig.~\ref{fig:TNKj}.
The holographic duals were discussed in \cite{Chaney:2018gjc}.
A quiver deformation for $N>jK$ is given by
\begin{align}\label{eq:TNKj-quiver-1}
[N-jK]-(N-jK+j-1)\stackrel{x_1}{-}
\ldots\stackrel{x_{K-1}}{-}(N&-K)\stackrel{x_K}{-}
\ldots \stackrel{x_{N-3}}{-}(2)-[2] ~,\nonumber \\
&\ \,| \, \\
&\,[j] \nonumber
\end{align}
with all Chern-Simons levels zero.
Between the links labeled by $x_{K}$ and $x_{N-3}$ the rank of the gauge groups decreases in steps of one.
There are $K-2$ gauge nodes between the links labeled by $x_1$ and $x_{K-1}$, with rank increasing in steps of $j-1$. 
For $j=1$ there is a total of $K$ $SU(N-K)$ gauge nodes. 
For $j=2$ and $N=2K$ the would-be $(1)$ gauge node on the left end is replaced by two fundamental hypermultiplets; this case was discussed in sec.~\ref{sec:T2K-K-2}.
The $T_{N,N-2,1}$ theories correspond to the $R_{0,N}$ theories of \cite{Chacaltana:2010ks,Bergman:2014kza}, and the $\chi_N^k$ theories of \cite{Bergman:2014kza} correspond to $T_{N,N-k-1,1}$.
The supergravity limit corresponds to $N,K\gg1$ with $j$ of order one.

In (\ref{eq:cF-gen-3}) the quiver (\ref{eq:TNKj-quiver-1}) corresponds to $L=N-2$, $N_t= N-jK+t(j-1)$ for $t\leq K$ and $N_t=N-t$ for $t>K$, as well as $k_1=N-jK$, $k_K=j$ and $k_{N-2}=2$. The continuous version in (\ref{eq:cF-gen-6}) is defined by
\begin{align}\label{eq:TNKj-data}
 N(z)&=N\begin{cases}
        1-j\mathds{k}+(j-1)z~, & z\leq\mathds{k}\\
        1-z~, & z\geq\mathds{k}
       \end{cases}~,
 &
 \mathds{k}&\equiv\frac{K}{N}~,
 \nonumber\\
 k(z)&=\frac{N-jK}{L}\delta(z)+\frac{j}{L}\delta(z-\mathds{k})+\frac{2}{L}\delta(z-1)~, & c(z)&=0~.
\end{align}
The quiver deformation has $N_f=2N$ at all interior nodes, such that $\mu(z)=\tau(z)=0$.
Explicitly, (\ref{eq:cF-gen-6}) becomes
\begin{align}\label{eq:cF-TNKj}
 \cF_{T_{N,K,j}} \ = \ &
 N^2\int_0^1 dz \int dx\,dy\,\cL
 +N^3j(N-K)\int dx\, \hat\rho(\mathds{k},x)F_H(x)
 \nonumber \\ &
+N^3(N-jK)^2 \int dx\, \hat\rho(0,x)\left[F_H(x)-\frac{1}{2}\int dy\, \hat\rho(0,y)F_H(x-y)\right]~. 
\end{align}
The flavors at the $SU(2)$ node on the right end only produce contributions that are subleading at large $N$, but the $j$ flavors at $z=\mathds{k}$ are important.
Since $N_f=2N$ at all internal nodes, the discussion of sec.~\ref{sec:Nf2NC-sol} applies.
The saddle point eigenvalue distribution is given by (\ref{eq:varrho-s}) with (\ref{eq:TNKj-data}), 
\begin{align}\label{eq:TNKj-rho}
 \hat\rho_{s}(z,x)&=
 \frac{1}{N(z)} \left(\frac{(N-jK) \sin (\pi  z)}{\cosh \left(2 \pi  x\right)-\cos (\pi  z)}
 -\frac{Nj}{2\pi}
   \ln \left(\frac{\cosh \left(2 \pi  x\right)-\cos (\pi  (\mathds{k}-z))}{\cosh \left(2 \pi  x\right)-\cos
   (\pi  (\mathds{k}+z))}\right)\right)~.
\end{align}
The free energy is given by (\ref{eq:free-energy-gen}) with (\ref{eq:cF-gen-Nf2NC}) and (\ref{eq:TNKj-data}), 
\begin{align}\label{eq:FS5-TNKj}
 F_{\vec{\omega}}^{T_{N,K,j}}&=
 -\frac{N^4}{8\pi^2}\frac{\omega_{\rm tot}^3}{\omega_1\omega_2\omega_3}\Bigg[
 (1-j\mathds{k})^2\zeta (3)
 +\frac{2j}{\pi }(1-j\mathds{k})D_4\big(e^{i \mathds{k} \pi }\big)
 + \frac{j^2}{2\pi ^2}\left(\zeta (5)-D_5\big(e^{2 i \mathds{k} \pi }\right)
 \Bigg],
\end{align}
with $D_n(z)$ defined in (\ref{eq:D-n-def}).
The limit $K\rightarrow 1$, leads back to the unconstrained $T_N$ theory, and (\ref{eq:FS5-TNKj}) reduces to (\ref{eqn:cFTN-eval-4}).
The limit $j\mathds{k}\rightarrow 1$ leads to the $T_{jK,K,j}$ theories, and for $j=2$ the result agrees with (\ref{eq:Tj2-result}). 
The result in (\ref{eq:FS5-TNKj}) shows that the parameters of the field theory can in general appear as arguments of polylogarithms, and that the form of the free energy is not limited to the simple dependence on the field theory parameters found in the examples of the previous sections. 

For the special cases studied numerically in \cite{Fluder:2019szh}, $F_{\vec{\omega}}^{T_{N,K,j}}$ in (\ref{eq:FS5-TNKj}) reduces to
\begin{align}
 F_{S^5}^{T_{N,N/2,1}}&=-\frac{27N^4}{32\pi^2}\left(\zeta (3)+\frac{4}{\pi}D_4(i)+\frac{31 }{8 \pi ^2}\zeta (5)\right)~,
 \nonumber\\
 F_{S^5}^{T_{N,N/4,3}}&=
 -\frac{27N^4}{32\pi^2}\left(
 \frac{\zeta(3)}{4}
 +\frac{6}{\pi }D_4\big(e^{\frac{i\pi}{4}}\big)
 +\frac{4743}{256 \pi ^2}\zeta (5)\right)~,
\end{align}
and matches the numerical field theory and supergravity results.
For general $j$, $\mathds{k}$ one can compare to the supergravity results of \cite{Fluder:2019szh},
by rearranging (\ref{eq:FS5-TNKj}) to match the parametrization of the free energy in (3.26) of \cite{Fluder:2019szh} as follows, 
\begin{align}
 F_{S^5}^{T_{N,K,j}}&=-\frac{27}{8 \pi ^2}\zeta(3)N^4\left[1-2j F_{T_{N,K,j}}^{(1)}+j^2F_{T_{N,K,j}}^{(2)}\right]~,
 \nonumber\\
 F_{T_{N,K,j}}^{(2)}&=2\mathds{k}F_{T_{N,K,j}}^{(1)}-\mathds{k}^2
 -\frac{D_5\left(e^{2 i \mathds{k} \pi }\right)-\zeta (5)}{2 \pi ^2 \zeta (3)}~,
 &
 F_{T_{N,K,j}}^{(1)}&=\mathds{k}-\frac{D_4\left(e^{i \mathds{k} \pi }\right)}{\pi\zeta (3)} ~.
\end{align}
With this parametrization $F_{T_{N,K,j}}^{(1)}$ matches the plot of $\mathcal S^{(1)}$ in \cite{Fluder:2019szh} and $F_{T_{N,K,j}}^{(2)}$ matches $\mathcal S^{(2)}$.

\subsection{\texorpdfstring{$+_{N,M,j}$}{+[N,M,j]} theory}

The brane realization of the $+_{N,M,j}$ theories is obtained from that of the $+_{N,M}$ theories by partitioning one group of $N$ D5-branes into $N/j$ subgroups and terminating each subgroup on a single D7-brane, as shown in fig.~\ref{fig:plus-NMk}. 
These theories describe the UV fixed points of the quiver gauge theories
\begin{align}\label{eq:plus-NMj-quiver}
 (j)-(2j)-\ldots - (N-2j) - (N-j) - &(N) - (N)^{N-\frac{N}{j}-1}-[N]~.
 \nonumber\\
&\;\ | \\
&\;[j]& \nonumber 
\end{align}
For $j=1$ the would-be $(1)$ gauge node on the left end is replaced by two fundamental hypermultiplets, $[2]$. The Chern-Simons levels are zero for all nodes.
The supergravity limit corresponds to $N,M\gg1$ and $j$ of order one.
Aspects of the spectrum were studied in \cite{Bergman:2018hin}.

In (\ref{eq:cF-gen-3}) the quiver for the $+_{N,M,j}$ theories with $j>1$ corresponds to $L=M-1$, 
$N_t=tj$ for $t\leq N/j$ and $N_t=N$ for $t\geq N/j$, with $k_{N/j}=j$ and $k_L=N$.
For $j=1$ the first gauge node is replaced by 2 fundamental flavors.
In the large-$N$ limit the continuous version of $\cF_{+_{N,M,j}}$ is given by (\ref{eq:cF-gen-6}) with
\begin{align}\label{eq:plus-NMk-data}
 N(z)&=\begin{cases}
         Mjz~, &z\leq \mathds{k}\\
         N~,& z\geq \mathds{k}
       \end{cases}\,,
 &
 \mathds{k}&\equiv\frac{N}{jM}~,
\end{align}
and
\begin{align}
k(z)&=\frac{2}{L}\delta_{j,1}\delta(z)+\frac{j}{L}\delta(z-\mathds{k})+\frac{N}{L}\delta(z-1)~,
&
 c(z)&=0~.
\end{align}
Since $N_f=2N$ at all interior nodes, $\mu(z)=\tau(z)=0$.
Explicitly, (\ref{eq:cF-gen-6}) becomes
\begin{align}\label{eq:cF-plus-NMk}
  \cF_{+_{N,M,j}} \ = \ &
 M^2\int_0^1 dz \int dx\,dy\,\cL
 -\frac{1}{2}M^3N^2\int dx\,dy\, \hat\rho(1,x)\hat\rho(1,y)F_H\big(x-y\big)
 \nonumber\\ & 
 +M^3Nj\int dx\, \hat\rho(\mathds{k},x)F_H(x)
 +M^3N^2\int dx\,\hat\rho(1,x)F_H(x)
 ~.
\end{align}
As in previous examples, the flavors at the left end of the quiver that appear for $j=1$ produce only subleading contributions and drop out.

Since $N_f=2N$ at all interior nodes, the saddle point eigenvalue distributions are given by (\ref{eq:varrho-s}) with (\ref{eq:plus-NMk-data}), 
\begin{align}\label{eq:plus-NMk-saddle}
\hat\rho_s(z,x)&=
\frac{1}{N(z)} \left(
\frac{N \sin (\pi  z)}{\cosh \left(2 \pi x\right)+\cos (\pi  z)}
+
\frac{j M}{2\pi}\ln \left(\frac{\cosh \left(2 \pi  x\right)-\cos (\pi  (\mathds{k}+z))}{\cosh \left(2 \pi  x\right)-\cos (\pi(\mathds{k}-z))}\right)\right)~.
\end{align}
The free energy is given by (\ref{eq:free-energy-gen}) with (\ref{eq:cF-gen-Nf2NC}),
\begin{align}\label{eq:FS5-plus-NMk}
 F_{\vec{\omega}}^{+_{N,M,j}}&=
  -\frac{M^2}{8\pi^2}\frac{\omega_{\rm tot}^3}{\omega_1\omega_2\omega_3}\Bigg[
  N^2\zeta(3)
  +\frac{2jM N}{\pi}D_4\big(e^{i \pi(1- \mathds{k})}\big)
  +\frac{j^2M^2}{2\pi^2} \left(\zeta(5)-D_5\big(e^{2 i\mathds{k} \pi }\big)\right)  \Bigg],
\end{align}
with $D_n$ defined in (\ref{eq:D-n-def}). 
For $j=1$ and $M>N$, the brane web for the $+_{N,M,j}$ theory is equivalent (after moving 7-branes) to the one for the $T_{M,M-N,1}$ theory.
The free energy indeed agrees with (\ref{eq:FS5-TNKj}) after identifying the parameters appropriately.
Strictly speaking, the case $j=N$, leading back to the unconstrained $+_{N,M}$ theory, is outside the range of validity for this result, since $j$ was assumed to be of order one for the derivation. By an appropriate expansion for small $\mathds{k}$, however, one can still recover (\ref{eqn:cFMN-eval-4}).

\section{Discussion}\label{sec:discussion}

We have obtained exact results for the free energies of 5d SCFTs with holographic duals in Type IIB on squashed spheres.
The SCFTs have relevant deformations that flow to quiver gauge theories of the form (\ref{eq:quiver}), and in the large-$N$ limits described by supergravity the number of quiver nodes is large. The ranks of at least some gauge groups are large as well, but not necessarily all of them. 
The saddle point conditions for the matrix models resulting from supersymmetric localization of the squashed sphere partition function were formulated as problems akin to 2d electrostatics, which can be solved using standard methods.
They were solved explicitly for theories with $N_f=2N$ at all interior gauge nodes, and for a sample of theories with $N_f\neq 2N$ nodes including theories with Chern-Simons terms.
The conformal central charge at large $N$ was shown to generally be related to the round sphere free energy by $C_T=-640\pi^{-2}F_{S^5}$.

For a number of theories the saddle point solutions and free energies were discussed explicitly.
This includes the $+_{N,M}$, $T_N$, $Y_N$ and $\pslash_N$ theories, which are engineered by the unconstrained 5-brane junctions in fig.~\ref{fig:unconstrained}.
They admit quiver gauge theory deformations given in (\ref{eq:D5NS5-quiver}), (\ref{TNquiver}), (\ref{eq:YN-quiver-2}) and (\ref{eq:pslash-quiver}), respectively. 
These theories do not have fundamental flavors at internal nodes.
The free energies on squashed spheres with metric (\ref{eq:squashed-metric}), where $\phi_i$ are the squashing parameters, are
\begin{align}
 F_{\vec{\omega}}^{+_{N,M}}&=-\frac{7}{16\pi^2}\frac{\omega_{\rm tot}^3}{\omega_1\omega_2\omega_3}\zeta(3)N^2M^2~,
 &
 F_{\vec{\omega}}^{T_N}&=-\frac{1}{8\pi^2}\frac{\omega_{\rm tot}^3}{\omega_1\omega_2\omega_3}\zeta(3)N^4~,
 \nonumber\\
 F_{\vec{\omega}}^{Y_N}&=-\frac{1}{2\pi^2}\frac{\omega_{\rm tot}^3}{\omega_1\omega_2\omega_3}\zeta(3)N^4~,
 &
 F_{\vec{\omega}}^{\pslash_N}&=-\frac{7}{2 \pi ^2} \frac{\omega_{\rm tot}^3}{\omega_1\omega_2\omega_3}\zeta (3) N^4~.
 \label{eq:unconstrained-summary}
\end{align}
The squashing parameters are encoded in $\omega_i=1+i\phi_i$, with $\omega_{\rm tot}\equiv\omega_1+\omega_2+\omega_3$,
and the round $S^5$ is recovered for $\omega_1=\omega_2=\omega_3=1$.
We also discussed the $T_{2K,K,2}$, $T_{N,K,j}$ and $+_{N,M,k}$ theories, which are engineered in Type IIB 
by the constrained 5-brane junctions shown in fig.~\ref{fig:constrained}.
Their quiver gauge theory deformations are given in (\ref{eq:TNKj-quiver-2}), (\ref{eq:TNKj-quiver-1}) and (\ref{eq:plus-NMj-quiver}), respectively, and these have fundamental flavors at interior nodes. The free energies are
\begin{align}\label{eq:constrained-summary}
 F_{\vec{\omega}}^{T_{N,K,j}}&=
 -\frac{N^4}{8\pi^2}\frac{\omega_{\rm tot}^3}{\omega_1\omega_2\omega_3}\Bigg[
 (1-j\mathds{k})^2\zeta (3)
 +\frac{2j}{\pi }(1-j\mathds{k})D_4\big(e^{i\pi \mathds{k}  }\big)
 + \frac{j^2}{2\pi ^2}\left(\zeta (5)-D_5\big(e^{2 i\pi \mathds{k}  }\right)
 \Bigg],
 \nonumber\\
 F_{\vec{\omega}}^{+_{N,M,j}}&=
  -\frac{M^2}{8\pi^2}\frac{\omega_{\rm tot}^3}{\omega_1\omega_2\omega_3}\Bigg[
  N^2\zeta(3)
  +\frac{2jM N}{\pi}D_4\big(e^{i \pi(1- \mathds{k})}\big)
  +\frac{j^2M^2}{2\pi^2} \left(\zeta(5)-D_5\big(e^{2 i\pi\mathds{k} }\big)\right)  \Bigg],
\end{align}
where $D_4(e^{i\alpha})=\Im({\rm Li}_4(e^{i\alpha}))$ and $D_5(e^{i\alpha})=\Re({\rm Li}_5(e^{i\alpha}))$.
For the  $T_{N,K,j}$ theory $\mathds{k}=K/N$, and for the $+_{N,M,j}$ theory $\mathds{k}=N/(jM)$.
The $T_{2K,K,2}$ theory was discussed separately since the gauge theory deformation involves extra flavors, 
but the free energy is given by setting $N=2K$, $j=2$ in the result for the generic $T_{N,K,j}$ theories.
The results for theories with flavors at interior nodes involve polylogarithms up to degree five, while the maximal degree for theories without flavors is three.\footnote{%
However, taking inspiration from  \cite{DHoker:2019blr}, where transcendentality weight $n$ is assigned to $\zeta(n)$ with $n\geq 2$, such that $\pi$ has weight one, and extending it so as to assign weight $n$ to ${\rm Li}_n(z)$ for an arbitrary phase $z$, 
gives a homogeneous transcendentality weight for the free energies of all theories considered.}
Moreover, while the results in (\ref{eq:unconstrained-summary}) show a simple dependence on the field theory parameters, in (\ref{eq:constrained-summary}) the field theory parameters appear as arguments of polylogarithms.

The analytic results for the $S^5$ free energies of the $+_{N,M}$ and $T_N$ theories match the analytic supergravity results of \cite{Gutperle:2017tjo} and the numerical field theory computations of \cite{Fluder:2018chf}. 
For the $T_{N,K,j}$ theories the $S^5$ free energy matches the numerical supergravity and field theory results of \cite{Fluder:2019szh}.
For the $Y_N$ and $\pslash_N$ theories the supergravity results are included in \cite{Gutperle:2017tjo}, and match the analytic field theory results (\ref{eq:unconstrained-summary}). Finally, for the $+_{N,M,k}$ theories a supergravity computation for a sample of parameter choices matches the analytic result in (\ref{eq:constrained-summary}).
The results presented here therefore support the AdS$_6$/CFT$_5$ dualities proposed in \cite{DHoker:2016ysh,DHoker:2017mds} for 5d SCFTs engineered by 5-brane junctions, and the dualities proposed in \cite{Gutperle:2018vdd,Bergman:2018hin,Chaney:2018gjc} for AdS$_6$ solutions with 7-branes \cite{DHoker:2017zwj}.

The Type IIB supergravity description for the 5d SCFTs considered here on a family of squashed spheres with $-\phi_1=\phi_2=\phi_3=i\sqrt{1-s^2}$ can be obtained with the consistent truncations from the Type IIB solutions of \cite{DHoker:2016ysh,DHoker:2017mds,DHoker:2017zwj} to 6d gauged supergravity worked out in \cite{Hong:2018amk,Malek:2018zcz,Malek:2019ucd}. Namely, by uplifting the 6d solutions of \cite{Alday:2014rxa,Alday:2014bta}. 
The relation between the free energies on these squashed spheres and on the round $S^5$ found in \cite{Alday:2014rxa,Alday:2014bta}, $F_s=\frac{1}{27}(3-\sqrt{1-s^2})^3/(1-\sqrt{1-s^2})F_{S^5}$, therefore also holds when computing the free energies in Type IIB supergravity. This relation is a special case of the general relation (\ref{eq:FSomega-FS5-0}), providing a further match between field theory and supergravity.
In a similar vein, the universal relation between conformal central charge and sphere free energy derived in sec.~\ref{sec:CT} follows on the supergravity side from the existence of a consistent truncation to 6d supergravity, as spelled out in \cite{Fluder:2018chf}.

The analytic results obtained here provide a stepping stone for many more field theory studies. With the saddle point eigenvalue distributions in hand, one may compute other BPS quantities such as Wilson loops or supersymmetric correlators, or the flavor central charges from mass deformations following \cite{Chang:2017cdx}.
One may also expect similar methods to allow for an analytic computation of the topologically twisted indices studied numerically in \cite{Fluder:2019szh}, and matched to the supergravity prediction based on the solutions of \cite{Suh:2018tul,Suh:2018szn,Suh:2019ily}.
More generally, it would be interesting to study long quiver gauge theories in other dimensions, such as the 4d SCFTs discussed in \cite{Aharony:2012tz}.

\begin{acknowledgments}
It is a pleasure to thank Martin Fluder and Morteza Hosseini for fruitful collaborations and discussions,
and Diego Rodriguez-Gomez for interesting discussions. 
I am also grateful to the organizers and participants of the workshop ``Holography, Generalized Geometry and Duality'' at the Mainz Institute for Theoretical Physics for the inspiring workshop, and to MITP for hospitality.
This work is generously supported by the Mani L. Bhaumik Institute for Theoretical Physics.
\end{acknowledgments}

\appendix
\renewcommand\theequation{\Alph{section}.\arabic{equation}}

\section{Free energies for \texorpdfstring{$N_f=2N$}{Nf=2Nc} theories}\label{sec:cF-details}

In this appendix, $\cF$ in (\ref{eq:cF-gen-6}) is evaluated on the saddle point configuration for quivers with $N_f=2N$ at all interior nodes as given in (\ref{eq:hat-rho-varrho-s}) with (\ref{eq:varrho-s}).
This is done by directly evaluating $\cF$ in (\ref{eq:cF-gen-7}) on (\ref{eq:varrho-s}).
With the Chern-Simons terms and the Lagrange multiplier terms vanishing, (\ref{eq:cF-gen-7}) reduces to
\begin{align}\label{eq:cF-gen-eval}
 \cF\big\vert_{\varrho=\varrho_s} = \, &
 L^2\int_0^1 dz \int dx\,dy\, \cL\big\vert_{\varrho=\varrho_s}
 -\frac{1}{2}L^3\sum_{z\in\lbrace 0,1\rbrace}\int dx\,dy\, \varrho_s(z,x)\varrho_s(z,y)F_H\big(x-y\big)
 \nonumber\\ &
 +L^4\int_0^1dz\int dx\, \varrho_s(z,x)k(z)F_H(x)~.
\end{align}
With the boundary condition (\ref{eq:rho-bc-gen-rep}), the explicit boundary terms vanish, 
as do possible flavor contributions from the first and last node, leaving
\begin{align}\label{eq:cF-gen-eval-2}
 \cF = \, &
 L^2\int_0^1 dz \int dx\,dy\, \cL\big\vert_{\varrho=\varrho_s}
 +L^4\int_0^1dz\int dx\, \varrho_s(z,x)k(z)F_H(x)~.
\end{align}
With integration by parts, this becomes
\begin{align}
 \cF\big\vert_{\varrho=\varrho_s} = \, &
 -\frac{1}{2}L^2\int dx\,dy\,\left[\varrho_s(z,x)\partial_z\varrho_s(z,y)\right]_{z=0}^{z=1}F_H(x-y)
 \nonumber\\ &
 +L^2\int dx\,dy\,\varrho_s(z,x)\left(\frac{1}{8}\partial_x^2\varrho_s(z,x)+\frac{1}{2}\partial_z^2\varrho_s(z,x)\right)F_H(x-y)
 \nonumber\\ &
 +L^4\int_0^1dz\int dx\, \varrho_s(z,x)k(z)F_H(x)~.
\end{align}
The first term can be simplified using the boundary condition (\ref{eq:rho-bc-gen-rep}).
For the second term one uses the saddle point equation (\ref{eq:rho-eom-nf2nc}). 
With $k(z)$ in (\ref{eq:k-z-disc}) this leads to
\begin{align}
 \cF\big\vert_{\varrho=\varrho_s} = \, &
 -\frac{1}{2}L^2\int dx\,\left[N(z)\partial_z\varrho_s(z,x)\right]_{z=0}^{z=1}F_H(x)
 +\frac{1}{2}L^3\sum_{t=2}^{L-1}k_t \int dx\, \varrho_s(z_t,x)F_H(x)~.
\end{align}
The saddle point configurations (\ref{eq:varrho-s}) are symmetric under $x\rightarrow -x$, so the integral can be reduced to non-negative $x$ at the expense of a factor $2$, thus eliminating the absolute values in $F_H$. Explicitly,
\begin{align}\label{eq:cF-gen-eval-f}
 \cF\big\vert_{\varrho=\varrho_s} = \, &
 \frac{\pi}{6} L^2\int_0^\infty dx\,x^3\left[
 \left[N(z)\partial_z\varrho_s(z,x)\right]_{z=0}^{z=1}
 -L\sum_{t=2}^{L-1}k_t\varrho_s (z_t,x)\right]~.
\end{align}
This can be further evaluated using the explicit solution (\ref{eq:varrho-s}), leading to integrals that can be found e.g.\ in \cite{gradshteyn07} or performed using Mathematica.

To explicitly evaluate the integral in (\ref{eq:cF-gen-eval-f}), $\varrho_s$ is decomposed as follows,
\begin{align}
 \varrho_s(z,x)&=\varrho_0(z,x)-\frac{L}{2\pi}\sum_{t=1}^{L-1}k_t \tilde\varrho_t(z,x)~,
\nonumber\\
\varrho_0(z,x)&= \frac{N(0)\sin(\pi z)}{\cosh(2\pi x)-\cos(\pi z)}+\frac{N(1)\sin(\pi z)}{\cosh(2\pi x)+\cos(\pi z)}~,
\nonumber\\
\tilde\varrho_t(z,x)&= \ln \left(\frac{\cosh (2 \pi  x)-\cos\left(\pi(z-z_t)\right)}{\cosh (2 \pi  x)-\cos\left(\pi(z+z_t)\right)}\right)~.
\end{align}
Correspondingly splitting the terms in (\ref{eq:cF-gen-eval-f}) according to their power in $L$ leads to
\begin{align}\label{eq:cF-exp-app}
 \cF\big\vert_{\varrho=\varrho_s} &=L^2\cF_0+L^3\sum_{t=2}^{L-1}k_t\cF_t+L^4\sum_{t=2}^{L-1}\sum_{s=2}^{L-1}k_s k_t\cF_{st}~,
\end{align}
with $\cF_0$, $\cF_t$ and $\cF_{st}$ independent of $L$.
$\cF_0$ can be evaluated using integration by parts,
\begin{align}
\cF_0&=\frac{\pi}{6}\int_0^\infty dx\,x^3\left[N(z)\partial_z\varrho_0(z,x)\right]_{z=0}^{z=1}
\nonumber\\
&=
\frac{\pi}{6}\int_0^\infty dx\,x^3\left[\frac{(N(0)+N(1))^2}{2}\partial_x\csch(2\pi x)+\frac{(N(0)-N(1))^2}{2}\partial_x(\coth(2\pi x)-1)\right]
\nonumber\\
&=-\frac{2 N(0)^2+3 N(0) N(1)+2 N(1)^2}{16 \pi ^2}\zeta (3)~.
\label{eq:cF0-app}
\end{align}
Evaluating $\cF_t$ leads to
\begin{align}
 \cF_t&=-\frac{\pi}{6}\int_0^\infty dx\,x^3\left[\varrho_0(z_t,x)+\frac{1}{2\pi}\left[N(z)\partial_z\tilde\varrho_t(z,x)\right]_{z=0}^{z=1}\right]
 \nonumber\\
 &=-\frac{\pi}{6}\int_0^\infty dx\,x^3\left[
 \frac{2 \sin (\pi  z_t)  N(0)}{\cosh (2 \pi  x)-\cos (\pi  z_t)}+\frac{2 \sin (\pi  z_t) N(1)}{\cosh (2 \pi x)+\cos (\pi z_t)}
 \right]
 \nonumber\\
 &=
 \frac{iN(0)}{8\pi^3}\left({\rm Li}_4\big(e^{i\pi z_t}\big)-{\rm Li}_4\big(e^{-i\pi z_t}\big)\right)
 +\frac{iN(1)}{8\pi^3}\left({\rm Li}_4\big(e^{i\pi (1-z_t)}\big)-{\rm Li}_4\big(e^{-i\pi (1-z_t)}\big)\right)~.
 \label{eq:cF1-app}
\end{align}
Finally,
\begin{align}
\cF_{st}&=\frac{1}{12}\int_0^\infty dx\,x^3\tilde\varrho_s(z_t,x)
=-\frac{1}{48}\int_0^\infty dx\,x^4\partial_x\tilde\varrho_s(z_t,x)
\nonumber\\
&=
-\frac{\pi}{12}\int_0^\infty dx\,x^4
\frac{\sinh (2 \pi  x) \sin (\pi z_s) \sin (\pi z_t)}{(\cosh (2 \pi  x)-\cos (\pi  (z_s-z_t)))   (\cosh (2 \pi  x)-\cos (\pi  (z_s+z_t)))}
\nonumber\\
&=\frac{1}{32\pi^4}\left(
{\rm Li}_5\big(e^{i\pi(z_s+z_t)}\big)+{\rm Li}_5\big(e^{-i\pi(z_s+z_t)}\big)
-{\rm Li}_5\big(e^{i\pi(z_s-z_t)}\big)-{\rm Li}_5\big(e^{-i\pi(z_s-z_t)}\big)
\right)~.
\label{eq:cF2-app}
\end{align}
The combination of (\ref{eq:cF-exp-app}) with (\ref{eq:cF0-app}), (\ref{eq:cF1-app}) and (\ref{eq:cF2-app}) establishes (\ref{eq:cF-gen-Nf2NC}).

\bibliography{FS5}
\end{document}